\documentclass[manuscript]{acmart}
\AtBeginDocument{%
  }

\setcopyright{none}
\copyrightyear{2018}
\acmYear{2018}
\acmDOI{XXXXXXX.XXXXXXX}
\acmConference[Conference acronym 'XX]{Make sure to enter the correct
  conference title from your rights confirmation email}{June 03--05,
  2018}{Woodstock, NY}
\acmISBN{978-1-4503-XXXX-X/2018/06}

\usepackage{multicol}
\usepackage{multirow}
\usepackage{graphicx}
\begin{document}

\title{Slide to Report: Continuous Reporting Symptom-Specific Cybersickness in VR}

\author{Purnata Saha}
\email{purnata.saha@my.utsa.edu}
\affiliation{%
  \institution{University of Texas at San Antonio}
  \city{San Antonio}
  \state{Texas}
  \country{USA}
}

\author{Rifatul Islam}
\email{rislam11@kennesaw.edu}
\affiliation{%
  \institution{Kennesaw State University}
  \city{Kennesaw}
  \state{Georgia}
  \country{USA}
}
\author{Qiliang He}
\email{qiliang.he@utsa.edu}
\affiliation{%
  \department{Psychology}
  \institution{University of Texas at San Antonio}
  \city{San Antonio}
  \state{Texas}
  \country{USA}
}
\author{John Quarles}
\email{john.quarles@utsa.edu}
\affiliation{%
  \institution{University of Texas at San Antonio}
  \city{San Antonio}
  \state{Texas}
  \country{USA}
}

\renewcommand{\shortauthors}{Saha et al.}

\renewcommand{\shortauthors}{Saha et al.}
\renewcommand{\shortauthors}{Saha et al.}

\begin{abstract}
Cybersickness is often assessed during VR exposure through discrete overall sickness, which cannot capture specific symptoms or their changes over time. Post-exposure questionnaires such as the Simulator Sickness Questionnaire provide symptom-specific measures but cannot be collected during immersion. We investigate whether separating cybersickness into three symptom-specific dimensions—nausea, oculomotor discomfort, and disorientation—during immersion provides more informative continuous measurements. We conducted a between-subjects study with 52 participants in a VR-rollercoaster, comparing a three-slider interface with a single-slider interface. Concurrently,
participants also completed an auditory Simon task to measure attention. We additionally collected multimodal set of physiological signals, including GSR, ECG, PPG, eye and head tracking. Results show that each slider correlated most strongly with
its corresponding SSQ subscale, whereas the single slider correlated with all SSQ subscales. Using three sliders did not significantly
affect usability, workload, or attention. We publicly release the multimodal dataset to support future cybersickness research.

Our dataset and implementation are available in our \href{https://anonymous.4open.science/r/anonymized_data_multislider-8211/}{anonymized repository}.
\end{abstract}

\begin{teaserfigure}
  \centering
  \includegraphics[width=0.9\textwidth]{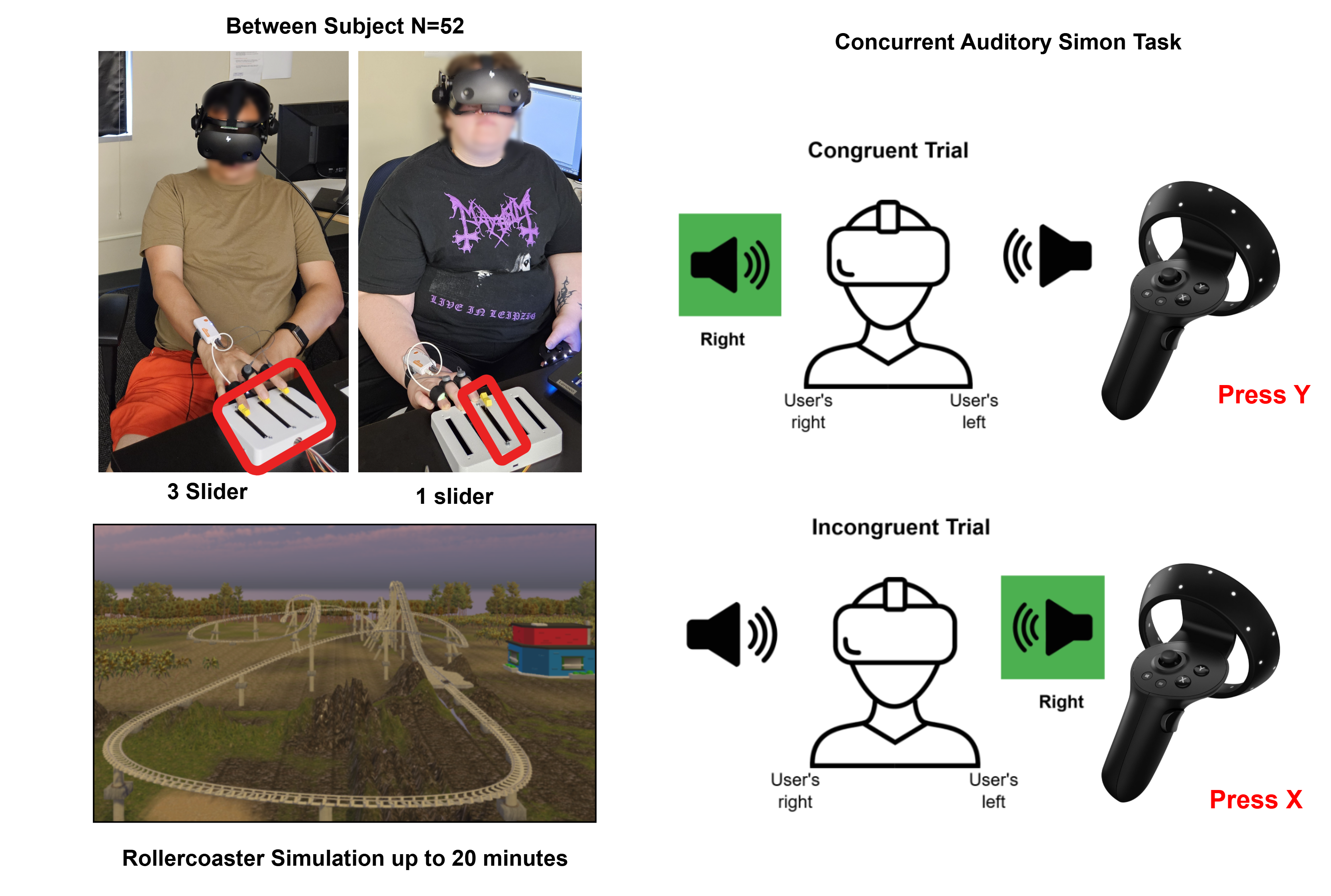}
\caption{Study overview. \textbf{Left, top:} Participants ($N = 52$,
between-subjects) were instrumented with an HP Reverb G2 Omnicept headset (eye
and head tracking), a Polar H10 chest strap for Electrocardiography (ECG), and a Shimmer Consensys unit
(Galvanic Skin Response(GSR) and Photoplethysmography (PPG), worn on the wrist and finger), and operated a 3D-printed physical
slider box with their dominant hand. Red outlines mark the active control in
each condition: three linear potentiometers mapped to nausea, disorientation,
and oculomotor discomfort in the three-slider condition (left), and the centre
potentiometer alone, reporting oculomotor discomfort, in the single-slider
condition (right). \textbf{Left, bottom:} The VR rollercoaster environment,
which participants experienced for up to 20 minutes or until they chose to
stop. \textbf{Right:} The concurrent auditory Simon task. Every 5\,s the spoken
word ``left'' or ``right'' was played in one ear, and participants responded
with the VR controller held in their non-dominant hand according to the
\emph{spatial location} of the sound ignoring its semantic content. In the
congruent example (top) the word ``right'' is played in the right ear,
requiring a right-ear response (Y); in the incongruent example (bottom) the
same word is played in the left ear, requiring a left-ear response (X).}
  \label{fig:teaser}
\end{teaserfigure}
\received{20 February 2007}
\received[revised]{12 March 2009}
\received[accepted]{5 June 2009}

\maketitle

\section{Introduction}
Cybersickness remains one of the main challenges in virtual reality. It is estimated that up to 80\% of VR users may experience symptoms of cybersickness after just 10 min of exposure to a virtual environment \cite{pawelczyk2025understanding,kim2021clinical}. The Simulator Sickness Questionnaire (SSQ) is the most commonly used tool for measuring cybersickness symptoms \cite{kennedy1993simulator}. It is designed to measure 3 subscales: nausea, disorientation, and oculomotor discomfort based on 16 defined questions of different symptoms \cite{walter2019brief}. However, SSQ is such a large questionnaire that it is typically administered before and/or after the simulation, not during immersion. As a result, the post-exposure SSQ scores may be influenced by recency bias, where participants tend to report symptoms based more on how they feel at the end of the experience rather than recalling the full progression of symptoms throughout the session \cite{lotfi2025recency}. During VR exposure, the Fast Motion Sickness scale (FMS) is commonly used, where participants report their discomfort on a 0–10 or 0–20 scale (e.g., 0 = no sickness, 10/20 = severe) at discrete intervals of 1–2 minutes \cite{keshavarz2011validating}. Participants are prompted via audio or visual cues to provide their feedback \cite{islam2020automatic, palmisano2023differences}. A limitation of FMS is that the term “discomfort” or "sickness" can be ambiguous, as participants may report general unease rather than symptoms specifically related to cybersickness \cite{reuten2021feelings}. The FMS score cannot uniquely identify the participant’s underlying cybersickness like SSQ \cite{keshavarz2011validating}. In addition, in simulations such as rollercoasters or racing scenarios, sickness can spike suddenly during sharp turns or drops, which may be missed by intermittent FMS \cite{cantory2025feasibility}. Continuous monitoring of FMS could capture these rapid changes more accurately. Cantory et al. \cite{cantory2025feasibility} investigated continuous self-reporting of FMS using a single 0–10 slider and reported that nearly 77\% of participants adjusted the slider multiple times within one minute. Although this approach improves the temporal resolution of self-reported sickness, it still captures only a single overall discomfort value and does not distinguish between different types of cybersickness symptoms.
Incorporating the SSQ subscales nausea, oculomotor, and disorientation into a continuous reporting system could allow participants to provide more specific and symptom focused feedback. In this study, we introduce a novel method in which participants continuously report their nausea, oculomotor, and disorientation based discomfort using three sliders (Linear potentiometer) placed in a 3D-printed box, enabling more precise and dynamic tracking of cybersickness during VR immersion. 
Beyond physical discomfort, cybersickness also influences cognitive processes such as attention, memory, and decision-making \cite{sokolowska2023impact}. Prior research suggests that cybersickness can divert attentional resources, leading to slower reaction times and reduced task accuracy in VR \cite{mimnaugh2023virtual, sepich2022impact}. Monitoring attention alongside cybersickness can therefore provide additional insight into how symptom severity correlates to attention or reaction time. To investigate this, we incorporate a cognitive dual-task paradigm during VR exposure. As the secondary task, we include an auditory Simon task, which measures attention and inhibitory control. In this task, participants respond based on which ear (left or right) they hear the sound in, while ignoring what the sound actually is \cite{burgoyne2024modality}.
We then compared performance across the single- and three-slider conditions to examine whether tracking one versus three symptoms affected participants’ reaction time, accuracy, or attention. This comparison will help evaluate the usability trade-offs of continuous, multidimensional cybersickness monitoring and support the design of more effective real-time assessment methods for future VR systems. 

In this study, we introduce a custom three-slider physical interface that enables users to dynamically and independently report specific cybersickness subscales - nausea, oculomotor, and disorientation, during VR immersion. Using this interface, we investigate the following two research questions:

\begin{itemize}
    \item \textbf{RQ1:} Does continuous, SSQ subscale-specific slider reporting correlate with established SSQ subscale scores?
   
   \item \textbf{RQ2:} Does using three sliders instead of one make the reporting task harder, in terms of reaction time, accuracy, and perceived usability?
\end{itemize}

\section{Related Work}
\subsection{Cybersickness assessment method}
Traditional cybersickness assessment methods can be broadly categorized into post-session questionnaires and during exposure assessment \cite{fernandez2025cybersickness, brown2021pre}. The most prominent of these is the SSQ, a 16-item scale rated from 0 to 3 (none to severe) that measures nausea, oculomotor, and disorientation subscales \cite{brown2022simulator}. Other variations include the Virtual Reality Sickness Questionnaire (VRSQ), a 9-item derivative of the SSQ focusing on oculomotor and disorientation factors \cite{kim2018virtual}; the Cybersickness Questionnaire (CSQ), which employs Likert-type ratings for VR-specific symptoms \cite{kourtesis2023cybersickness}. Among these, the SSQ remains the most widely used and reliable standard. However, while these tools are standardized and widely adopted, they are unable to capture symptom dynamics during immersion. 

To address this limitation, continuous assessment methods collect real-time subjective ratings during VR exposure. These typically utilize single scales, such as the Misery Scale (MISC) \cite{irmak2021objective}, which focuses on nausea using a 0--10 scale, or the FMS (0-10), which tracks overall discomfort \cite{keshavarz2011validating}. In traditional continuous assessments, participants are normally prompted to verbally report their discomfort scores at regular intervals of every few minutes \cite{davis2014systematic}.

\subsection{VR Datasets with Continuous Cybersickness} 
Most previous studies have relied on discrete FMS assessments during immersion, collected either verbally (audio reports) or visually (on-screen prompts requiring a click), rather than continuous measurement. However, in high-tension scenarios such as rollercoaster simulations \cite{islam2020automatic} or deliberately cybersickness-inducing environments \cite{zielasko2026not}, cybersickness symptoms can fluctuate rapidly, and continuous FMS monitoring may capture these changes more accurately than discrete sampling allows. To address this, some recent studies have adopted continuous, real-time self-report methods—such as sliders, dials, or joysticks—to capture moment-to-moment sickness severity throughout exposure.
Table~\ref{tab:continuous_studies_extended} summarizes recent studies using continuous self-report methods for cybersickness. Ryu and Kim compiled a publicly available dataset by having over 150 subjects use an external Microsoft Surface Dial to report their FMS score every 0.5 seconds during a virtual space flight \cite{ryu2026predicting}. Cantory et al. investigated the feasibility of continuous reporting by having 18 participants adjust a physical slider (a linear potentiometer) to report their sickness state during a virtual roller coaster ride \cite{cantory2025feasibility}. Their study demonstrated a statistically significant positive correlation between the continuously reported sickness ratings and simultaneously recorded tonic skin conductance levels. Plaza et al. compared three interfaces, bars, dials, and halos, each reporting three separate symptom dimensions: nausea, disorientation, and oculomotor discomfort. Tested with 25 participants using joystick input, all three matched CSQ-VR accuracy while capturing far more temporal detail.\\
Our study builds on this work by introducing a custom three-slider device, with each slider mapped to an SSQ subscale (nausea, oculomotor, disorientation). Like Plaza et al., we separate symptoms into distinct dimensions, but we use physical sliders instead of joystick input, giving participants proprioceptive feedback similar to Cantory et al's sliders but  we report subscale-specific values rather than one overall score. We also add a secondary auditory Simon task to measure attention during immersion, a dimension none of the prior studies test. By comparing user performance in a single-slider condition against our three-slider condition, we assess whether the added precision of multidimensional reporting can be achieved without significantly increasing cognitive load.

Finally, compared to existing datasets, our study captures a significantly more comprehensive array of multimodal data. Ryu and Kim collected only general FMS scores and virtual motion data, lacking physiological signals \cite{ryu2026predicting}. Cantory et al. reported one overall discomfort score paired exclusively with GSR \cite{cantory2025feasibility}. While Plaza et al. separated the three symptom subscales, they did not collect cardiovascular metrics like ECG or PPG \cite{plaza2026design}. In contrast, our study continuously records all three SSQ subscales paired with GSR, PPG, and ECG (during both baseline and simulation), alongside eye tracking, head tracking, and Simon task performance. This unparalleled combination of physiological, behavioral, and cognitive modalities allows us to directly relate specific symptom dimensions to autonomic arousal and attentional shifts within the same dataset.
\begin{table}[t]
\centering
\caption{Overview of Studies Utilizing Continuous Self-Report Methods for Cybersickness}
\label{tab:continuous_studies_extended}
\renewcommand{\arraystretch}{1.3}

\begin{tabular}{p{3.3cm} p{4.5cm} c p{4.7cm}}
\hline
\textbf{Study} & \textbf{Method} & \textbf{N} & \textbf{Modalities \& Availability} \\ \hline
Ryu and Kim et al (2026) \cite{ryu2026predicting} & Surface Dial (0--10), 2 Hz sampling rate & 150 & VR video, 6-DOF, head tracking. Public \\
\hline
Cantory et al (2025) \cite{cantory2025feasibility} & Linear potentiometer (4 Hz sampling rate) & 18 & GSR, Optical flow. Not public \\
\hline
Plaza et al (2026) \cite{plaza2026design} & Visual UIs: bars, dials, halos (joystick) & 25 & GSR, eye and head tracking. Not public dataset\\ \hline
\textbf{Our Study} & \textbf{3 physical linear potentiometers (4 Hz sampling rate)} & \textbf{52} & \textbf{Eye and head tracking, GSR, ECG, PPG, Auditory Simon task (Reaction time, Accuracy). Public} \\ \hline

\end{tabular}
\end{table}

\section{Experimental Design}
In designing this study, we focused on two primary decisions: selecting the devices for continuous, subscale-specific reporting, and choosing an appropriate secondary task to assess attentional differences between the single- and three-slider conditions.

\subsection{Interface Selection}
We began by reviewing prior literature on continuous cybersickness reporting devices, which have predominantly utilized linear potentiometers \cite{cantory2025feasibility}, dials \cite{ryu2026predicting}, or joysticks \cite{plaza2026design}. Cantory et al. tested several physical prototypes and found the linear slider to be the most intuitive and easiest to use \cite{cantory2025feasibility}. Similarly, Plaza et al.'s usability evaluation demonstrated that a visual bar-based interface—conceptually analogous to a linear slider—achieved higher usability scores than dial- or halo-based alternatives. Based on this empirical evidence, we selected the physical linear potentiometer for our interface, as it provides clear proprioceptive feedback, is highly usable, and remains cost-effective.
\subsection{Cognitive Load Assessment}
\begin{figure}[htbp]
\centering
\includegraphics[width=0.8\linewidth]{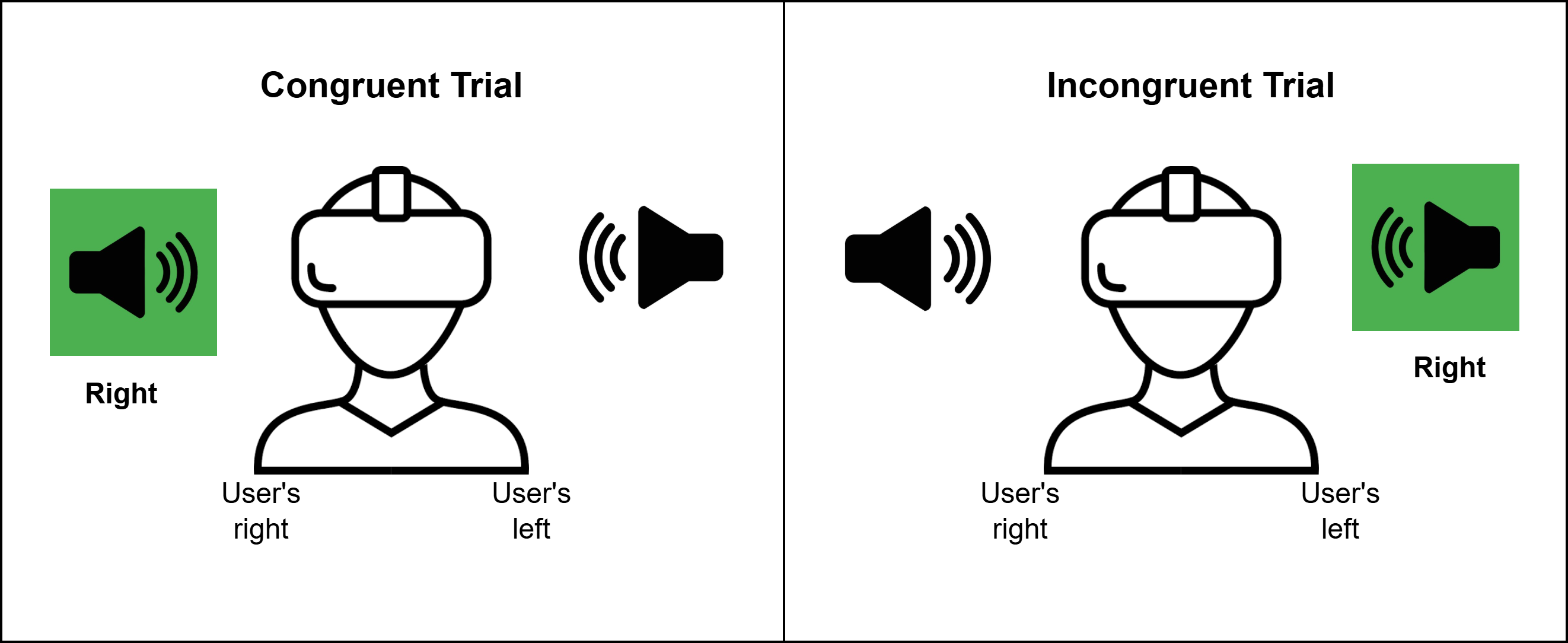}
\caption{Illustration of the auditory Simon task, adapted from Burgoyne et al. \cite{burgoyne2024modality}. Left: Example of a congruent trial. Right panel: example of a incongruent trial. The subject responds based on which ear received auditory input}
\label{fig:simon_task}
\end{figure}
To assess whether managing the three-slider interface imposed a higher cognitive load than the single-slider baseline, we implemented a cognitive dual-task paradigm \cite{burke2021multi}. An audio-based Simon task served as the secondary task during the VR simulation. The Simon task is a well-established behavioral measure of attentional control and cognitive conflict resolution \cite{manelis2023neural}. We selected this over more demanding cognitive assessments (e.g., Stroop \cite{chung2024understanding} or N-back tasks \cite{huang2025exploring, jaeggi2010concurrent}) to avoid overwhelming participants in an already highly stimulating environment. Furthermore, the N-back task imposes a substantial burden on working memory; consequently, poor performance can make it difficult to distinguish between generalized attentional depletion and inherent limitations in a participant's working memory \cite{owen2005n}. Specifically, we adapted the approach of Burgoyne et al., who developed an auditory Simon task in which participants must indicate the spatial location (which ear) of an auditory stimulus while actively suppressing its conflicting semantic content \cite{burgoyne2024modality}, illustrated in Figure~\ref{fig:simon_task}.

Participants operated the physical slider with their dominant hand and held a standard VR controller in their non-dominant hand to respond to the auditory cues. Every 5 seconds during the simulation, the spoken word "left" or "right" played in either the left or right ear. Participants (who all confirmed normal hearing) were instructed to respond solely to the sound's spatial location, completely ignoring its semantic meaning. The fixed response mapping utilized button X (left controller) or B (right controller) for left-ear sounds, and button Y (left controller) or A (right controller) for right-ear sounds. 
Trials were randomized between congruent (the word's meaning matched its spatial location) and incongruent (the meaning and location mismatched). Prior to the main simulation, participants completed a two-minute practice session in VR, viewing a plain grey screen to isolate the auditory task. To ensure task proficiency, any participant who made more than five errors received reiterated instructions and repeated the practice session.

\section{Material}
\subsection{Physical Slider}
\begin{figure}[htbp]
\centering
\includegraphics[width= 0.8\linewidth]{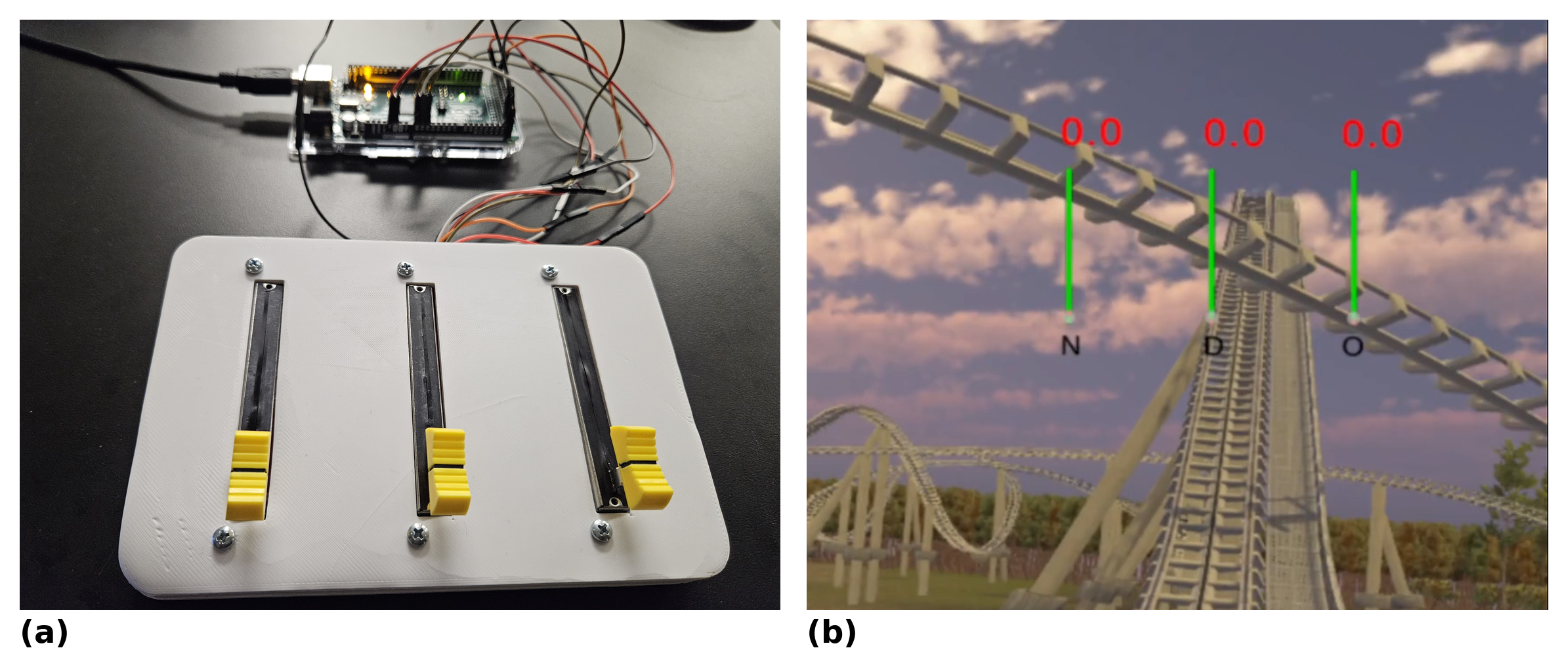}
\caption{The physical three-slider interface and corresponding in-VR visual feedback. (a) The physical apparatus, with the 3D-printed slider housing connected to the Arduino Mega. (b) The corresponding visual bars displayed in the VR environment, updating in real time as participants adjust each physical slider, mapped left to right to Nausea (N), Disorientation (D), and Oculomotor (O) discomfort.}
\label{fig:slider_setup}
\end{figure}

The 3D-printed housing was attached to the table in front of the participant using double-sided tape. As shown in Figure~\ref{fig:slider_setup}, the three-slider housing contained three linear potentiometers mounted with a 3 cm gap between each, while the single-slider housing used a single potentiometer mounted at the center. Each slider used a uxcell B103 10\,k$\Omega$ linear potentiometer (90 x 20\,mm), providing a dual analog output ranging from 0 to VCC.  Following the SSQ, we set the reporting scale from 0.0 to 3.0, allowing for continuous adjustments with a precision of 0.1. To provide visual feedback, we implemented corresponding visual bar sliders on a canvas within the Unity roller coaster environment, allowing participants to see their reported values change in real time as they adjusted each subscale.
For the three-slider condition, we followed the SSQ subscales because the SSQ is a widely used standard for measuring symptom-specific discomfort. The three sliders represented Nausea (N), Disorientation (D), and Oculomotor (O) discomfort, positioned from left to right, respectively. In the single-slider condition, the Oculomotor slider was positioned in the center, while the other two positions were left empty. Participants used the center slider to report only their Oculomotor symptoms.
An alternative design for the single-slider condition would have been to report overall discomfort using a single continuous FMS scale, comparable to Cantory et al \cite{cantory2025feasibility}. However, since the FMS scale (0--10) differs from the SSQ subscale scale (0.0--3.0) used in our three-slider condition, this would have introduced a scale mismatch, making a direct comparison between the single-slider and three-slider conditions unfair. We therefore retained the same 0.0--3.0 scale across both conditions, using only the Oculomotor subscale in the single-slider condition. The rationale for selecting Oculomotor discomfort specifically is described in the Pilot Study section.
\subsection{VR Hardware and Software}
The study used the HP Reverb G2 Omnicept Edition VR headset, which provides a display resolution of 2160 × 2160 pixels per eye and a refresh rate of 90\,Hz. The headset was used with its bundled motion controllers, each equipped with four built-in cameras for improved inside-out movement tracking. Audio was delivered through the headset's built-in headphones, and a Vive wireless adapter was used to enable untethered movement and ensure a smooth VR experience. The headset was connected to a Windows 10 computer with a 4.35\,GHz Intel Core i7 processor, 32\,GB of RAM, and an NVIDIA GeForce RTX 4080 graphics card. The virtual environment was developed in Unity 3D.
To capture eye and head tracking data, including gaze position and pupil size, we used the HP Omnicept SDK at a sampling rate of 120 Hz. This built-in eye-tracking capability of the Omnicept headset allowed us to collect gaze and pupillary data without additional external sensors.
\subsubsection{VR Environment} We used an open source roller coaster simulation, originally developed by Islam et al. \cite{islam2020automatic}. This simulation has been widely adopted in cybersickness research for its ability to reliably induce consistent motion cues. Each loop lasted approximately 57 seconds, followed by a 10-second pause before automatically transitioning into the next loop. An initial 20-second pause preceded the first loop, allowing participants time to acclimate before the ride began. Participants remained in the simulation for up to 20 minutes, unless they chose to stop early. While participants could freely adjust their head orientation to explore the surrounding environment, the roller coaster's trajectory and velocity were fixed and could not be influenced by the participant. The simulation ran continuously unless manually interrupted by the research team, and no handheld controllers were required for navigation.
\begin{figure*}[t]
    \centering
    \includegraphics[width=\linewidth]{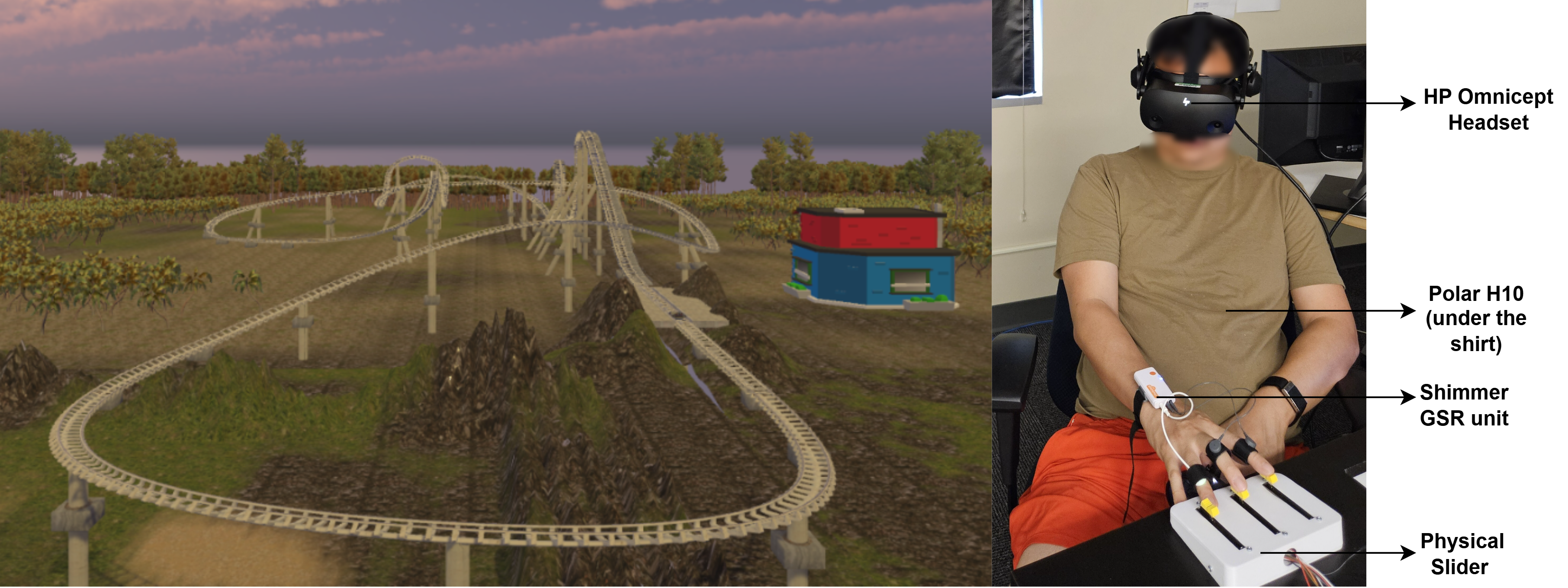}
    \caption{Overview of the VR environment and study apparatus. Left: The roller coaster
    simulation full track layout.  Right: A participant instrumented with the study's physiological sensors and
    subscale reporting slider: the HP Reverb G2 Omnicept headset (built-in eye and head
    tracking), a Polar H10 chest strap (worn under the shirt, can't be seen in the figure), a Shimmer Consensys GSR
    unit (GSR and PPG, worn on the wrist and finger), and the physical three-slider interface
    operated by the participant's dominant hand.}
    \label{fig:study-setup}

\end{figure*}
\subsection{Wearable Sensors}
For physiological data, we used a Shimmer Consensys GSR unit, worn on the participant's right hand. It captures GSR, PPG, and accelerometer and gyroscope data simultaneously. The GSR unit sampled at 128 Hz. ECG data were collected separately using a Polar H10 chest strap sensor, worn around the participant's chest. This sensor sampled at 120 Hz. Together, these sensors allowed us to continuously record GSR, PPG, ECG, accelerometer, and gyroscope data throughout both the baseline and simulation phases of the experiment.
Eye tracking data during the baseline period were collected at a sampling rate of 200 Hz using a Pupil Core eye tracker (Pupil Labs). All the modalities of collected data with the features and sampling rate is summarized in Table \ref{tab:data_modalities} in the appendix.

\section{User study}
\subsection{Pilot Study}
Prior to the main study, we conducted a pilot study ($N = 14$) employing a between-subjects design to prevent carryover effects \cite{doty2024cybersickness,nie2026revenge}. Participants were evenly divided, with seven evaluating the single-slider condition and seven evaluating the three-slider condition, both utilizing the same VR rollercoaster simulation planned for the main study. Guided by prior literature identifying nausea as a predominantly reported cybersickness symptom \cite{kim2022effects, garrido2022focusing}, we initially selected nausea as the representative metric for the single-slider pilot condition. However, analysis of the three-slider pilot data revealed that participants adjusted the oculomotor slider more frequently and reported significantly higher severity for oculomotor symptoms compared to nausea or disorientation. Consequently, we replaced nausea with oculomotor discomfort as the targeted symptom for the single-slider condition in the main study.

We adapted the continuous slider training protocol utilized by Cantory et al. \cite{cantory2025feasibility}. Participants viewed a desktop monitor positioned 1m away, which displayed the Nausea, Disorientation, and Oculomotor virtual sliders—identical to those used in the VR simulation—against a neutral grey background. After learning the specific physiological symptoms corresponding to each subscale, participants practiced adjusting the sliders across their 0.0 to 3.0 severity range to establish understanding of the physical-to-virtual interface mapping.

While Cantory et al. demonstrated that a five-minute session is sufficient for single-slider habituation, our pilot data indicated this was inadequate for a multi-dimensional interface. Specifically, the pilot showed inconsistent correlations between continuous in-VR reports and post-experience SSQ scores, which we attributed to participants struggling to distinguish between the three subscales. To address this, we increased the training duration from 5 to 10 minutes in the main study, ensuring participants were fully acclimated to both the symptom categorizations and the interface mechanics before beginning the simulation.

\subsection{Participants}
We conducted an a priori power analysis for our primary hypothesis, the
correlation between continuously reported slider values and their
corresponding $\Delta$SSQ subscale scores. Using G*Power, we
estimated the sample size required to detect a Pearson correlation of
$r = .5$--$.6$ with $\alpha = .05$ and power $= .80$, which indicated a
minimum of 20--30 participants per condition. 
We used a between-subjects design, rather than within-subjects, to avoid habituation and carryover effects that can arise when participants are repeatedly exposed to the same cybersickness-inducing simulation \cite{doty2024cybersickness}. A total of 52 participants (26 in each slider setup, $M_{age} = 28.3$, $SD = 7.9$, range 18--49) were recruited.. Table~\ref{tab:demographics} summarizes participant demographics and Figure~\ref{fig:age} shows the distribution of participant age, gender, and prior VR experience. The detailed study protocol is provided in Appendix~\ref{app:study_protocol}.

\begin{table}[h]
  \begin{minipage}{0.49\linewidth}
  \captionof{table}{Participant demographics (N = 52).}
    \centering
    \footnotesize
    \setlength{\tabcolsep}{4pt}
    \begin{tabular}{llr}
      \toprule
      \textbf{Category} & \textbf{Group} & \textbf{n (\%)} \\
      \midrule
      \multirow{3}{*}{Gender}
        & Male       & 29 (55.8\%) \\
        & Female     & 21 (40.4\%) \\
        & Non-binary & 2 (3.8\%) \\
      \midrule
      \multirow{3}{*}{Prior VR exp.}
        & Never        & 15 (28.8\%) \\
        & Rarely       & 29 (55.8\%) \\
        & Occasionally & 8 (15.4\%) \\
      \midrule
      \multirow{6}{*}{Ethnicity}
        & White / Caucasian & 15 (28.8\%) \\
        & Hispanic / Latino & 14 (26.9\%) \\
        & Asian             & 14 (26.9\%) \\
        & Multiracial/Other & 5 (9.6\%) \\
        & Middle Eastern    & 3 (5.8\%) \\
        & Black/Afr. Amer.  & 1 (1.9\%) \\
      \midrule
      \multirow{2}{*}{Corr. lenses}
        & Yes & 29 (55.8\%) \\
        & No  & 23 (44.2\%) \\
      \midrule
      \multirow{2}{*}{Conditions}
        & Motion sickness & 9 (17.3\%) \\
        & Migraines       & 5 (9.6\%) \\
      \bottomrule
      
    \end{tabular}
    
    \label{tab:demographics}
  \end{minipage}
  \hfill
  \begin{minipage}{0.50\linewidth}
    \centering
    \includegraphics[height=6.2cm,keepaspectratio]{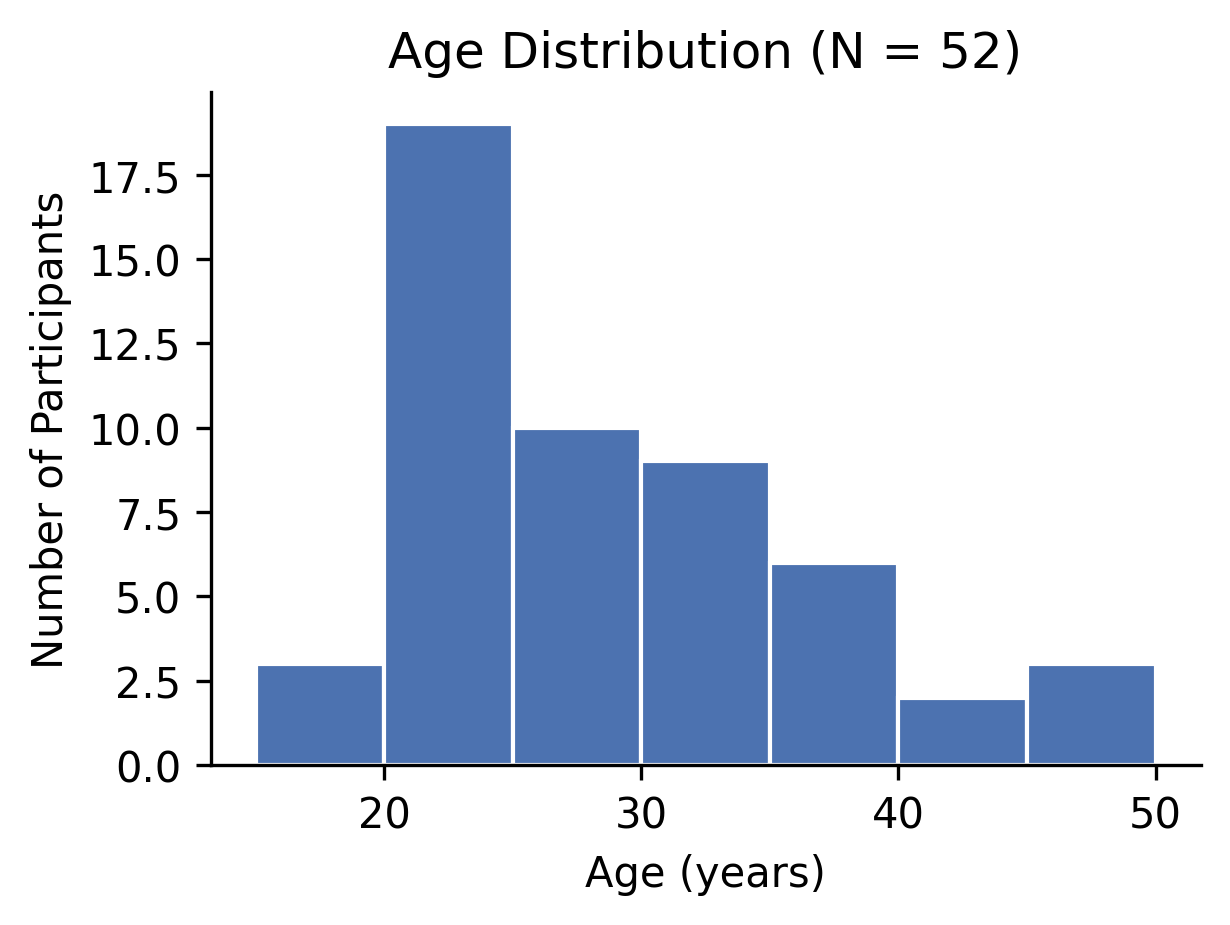}
    \captionof{figure}{Age distribution of participants (N = 52).}
    \label{fig:age}
    \Description{A histogram showing the age distribution of 52 study participants, ranging from 18 to 49 years, with the largest concentration in their early-to-mid twenties.}
  \end{minipage}
\end{table}

\subsection{Data Collection Procedure}
\begin{figure}[t]
    \centering
    \includegraphics[width=\linewidth]{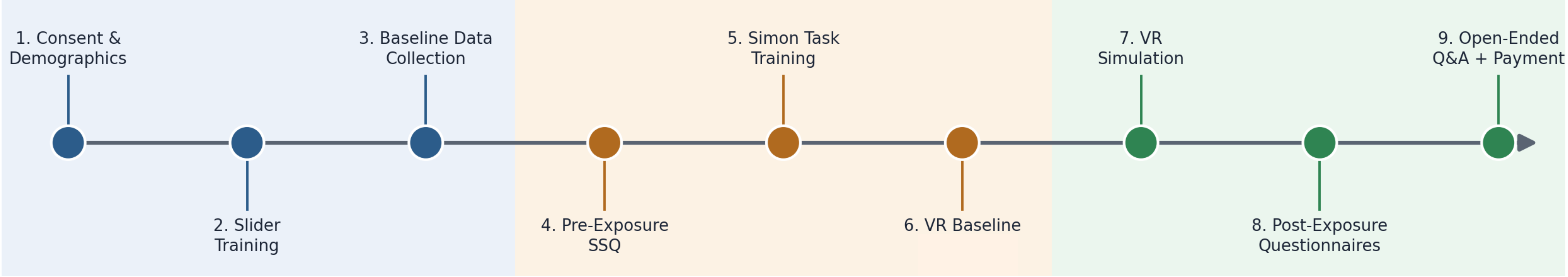}
    \caption{Timeline of the data collection procedure, grouped into three phases: preparation (steps 1--3), pre-immersion assessment (steps 4--6), and immersion \& debrief (steps 7--9).}
    \label{fig:data-collection-timeline}
\end{figure}
The summarized timeline of data collection is shown in Figure \ref{fig:data-collection-timeline}. Data collection proceeded through the following steps:

\begin{itemize}
    \item \textbf{Informed Consent and Demographics:} Participants provided informed consent and completed a demographic questionnaire.

\item \textbf{Slider Practice:} Participants completed a 10-minute training procedure tailored to their assigned condition (single-slider or three-slider). In the first 5 minutes, the experimenter explained which symptoms corresponded to the given subscale(s). Participants in the three-slider condition were informed about the inherent symptom overlap across the subscales; they were told that while all associated symptoms fall under each subscale (e.g., experiencing ``difficulty focusing'' would require increasing both the oculomotor and disorientation sliders). Participants in the single-slider condition were familiarized with Oculomotor symptoms and were instructed to adjust the slider based solely on these symptoms. They were informed that they might experience other symptoms during the VR session but were instructed not to use the slider to report symptoms that were not discussed during training. In the second 5 minutes, participants practiced mapping example scenarios to the correct slider(s) (e.g., ``If you started feeling dizzy at a slight level?''), with an opportunity to ask questions before data collection began. Each slider ranged from 0.0 (no discomfort) to 3.0, increasing with severity:
    \begin{itemize}
        \item \textit{Nausea:} Symptoms associated with this subscale included general discomfort, increased salivation, sweating, nausea, difficulty concentrating, stomach awareness, and burping. Participants were told to focus on the core symptoms first, particularly nausea, stomach awareness, increased salivation, and sweating. If they experienced symptoms that overlapped with another subscale, they were instructed to increase both corresponding sliders.
        \item \textit{Oculomotor (eyes):} This subscale includes general discomfort, fatigue, headache, eyestrain, difficulty focusing, difficulty concentrating, and blurred vision. Participants were instructed to interpret these symptoms as tired or heavy eyes, difficulty focusing or concentrating, or a headache behind the eyes. The core symptoms emphasized during training were eyestrain and headache.
        \item \textit{Disorientation (balance):} This subscale includes difficulty focusing, nausea, fullness of the head, blurred vision, dizziness (with eyes open or closed), and vertigo. Participants were trained to focus primarily on the core symptoms of dizziness, vertigo, and fullness of the head.
    \end{itemize}
    \item \textbf{Baseline Data Collection:} Baseline ECG, GSR, and PPG data were recorded using Polar and Shimmer devices while participants sat at rest without any external stimuli for 2 minutes. Participants were then instructed to look at the screen, and baseline eye-tracking data were collected for an additional 2 minutes while they viewed a plain grey screen with randomly appearing white calibration spheres. Participants were told to follow the spheres with their eyes. During this period, GSR and ECG data were recorded concurrently. The rationale for collecting the baseline data and for using the controlled grey-background, sphere  is described in Appendix~\ref{sec:appendix-baseline-rationale}.
   \item \textbf{Pre-SSQ:} Participants completed the SSQ prior to VR exposure.

    \item \textbf{Simon Task Training:} Before putting on the headset, participants were instructed to press the corresponding button with their non-dominant hand based on the location of the auditory stimulus.  Participants were informed that this served as their secondary task. While wearing the headset and viewing a plain grey screen displaying the task instructions, participants completed a 2-minute practice trial of the auditory Simon task. Participants who made more than five incorrect responses were provided with the instructions again and repeated the practice session.

   \item \textbf{VR Baseline:} Participants wore the VR headset for 2 minutes while viewing the same grey background like outside VR baseline with randomly appearing white calibration spheres in the virtual environment. They were instructed to visually follow the white spheres with their eyes as they appeared on the screen.

    \item \textbf{Simulation:} Participants began the roller coaster simulation, continuing until they chose to stop or until 20 minutes had elapsed. All participants concurrently performed the auditory Simon task. Of the 52 participants, 26 were assigned to the single-slider condition and 26 to the three-slider condition.

    \item \textbf{Post-Exposure Questionnaires:} Participants completed the post-exposure SSQ, the SUS to evaluate interface usability, and the NASA-TLX to assess perceived workload.

    \item \textbf{Open-Ended Questions and Payment:} Participants answered open-ended questions on whether the sliders were easy to use and which symptoms they felt most during the session. Participants were compensated \$30 per hour.
\end{itemize}

\subsection{Hypotheses}

Based on our research questions, we test the following hypotheses:

\begin{itemize}
    \item \textbf{H1:} In the three-slider condition, each continuously reported subscale (nausea, oculomotor, and disorientation) will be significantly correlated with its corresponding SSQ subscale score.
    \item \textbf{H2:} In the single-slider condition, the continuously reported oculomotor value will be significantly correlated with the SSQ oculomotor subscale score.
    \item \textbf{H3:} Simon reaction time and accuracy will be statistically equivalent between the
    three-slider and single-slider conditions
    \item \textbf{H4:} In both conditions, higher reported cybersickness will be associated with reduced attentional performance, reflected in slower reaction time and lower accuracy on the Simon task.
    \item \textbf{H5:} Perceived usability (SUS) and perceived workload (NASA-TLX)
    will be equivalent between the three-slider and single-slider conditions.
\end{itemize}
\section{Data Analysis}

This section describes the analyses conducted to address our research questions and hypotheses. We first describe data preprocessing, followed by analyses of slider usage patterns, relation with demographics, correlation between continuous subscale reports and SSQ scores (H1, H2), cognitive load between conditions (H3), the relationship between cybersickness and attention (H4) and usability (H5).

\subsection{Data Preprocessing}

Continuous slider values were logged throughout the simulation and synchronized with timestamps from the physiological, eye and head tracking recordings. For each participant, we computed the duration of time spent in the simulation, since participants could exit the roller coaster ride early. Nine of the 52 participants (17.3\%) ended the simulation before the 20-minute time limit, including four participants in the single-slider condition and five participants in the three-slider condition. All early exits occurred between 12 and 17 minutes after the start of the simulation. Physiological signals (GSR, PPG, ECG) were cropped according to the session start and end time.
We also extracted the eye and head tracking data for each participant.
For the Simon task, each trial's log recorded the trial type (congruent or incongruent), response correctness, and the time taken to press the corresponding button. From these logs, we computed reaction time (RT) and accuracy for each trial, which were used in subsequent analyses of cognitive load and attention.

\subsection{Slider Usage}

\begin{figure*}[htbp]
  \centering
  \includegraphics[width=\linewidth]{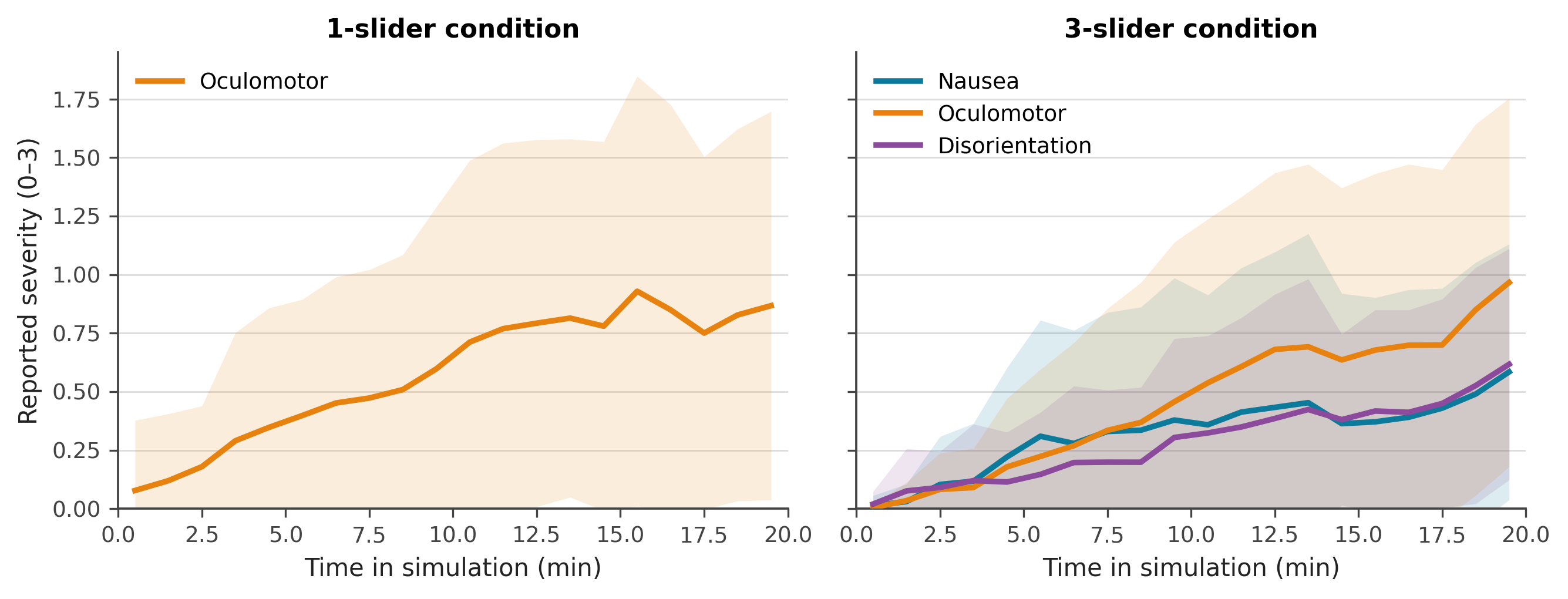}
  \caption{Reported symptom severity over the ride. (a) The single
  oculomotor slider in the 1-slider condition. (b) All three symptom axes
  in the 3-slider condition. Lines show the participant mean, shaded bands
  $\pm 1$\,SD.}
  \label{fig:timecourse}

\end{figure*}

Figure~\ref{fig:timecourse} shows how reported symptoms
changed over the ride averaged across
participants. We fitted a linear slope to each participant's reported severity over time.
Every slope was reliably positive: the single oculomotor slider rose by
$0.063$ units per minute ($p < .001$, $d = 0.95$), and in the three-slider
condition, nausea rose by $0.038$/min ($p = .004$, $d = 0.61$), oculomotor
by $0.051$/min ($p < .001$, $d = 0.83$), and disorientation by $0.035$/min
($p < .001$, $d = 0.75$). Symptoms built up at the same rate in both
conditions ($p = .28$, $r = .18$), so giving participants three sliders
instead of one did not change how fast they reported getting sick. 
In the three-slider condition, the three subscales climbed at
statistically indistinguishable rates (Friedman $p = .48$, $W = 0.03$).
Oculomotor discomfort ended highest ($M = 0.87$ in the final two minutes,
vs.\ $0.51$ for nausea and $0.54$ for disorientation).

Table~\ref{tab:usage} summarizes how participants in each condition engaged
with the reporting interface. Session duration and time-to-peak did not
differ between conditions (Mann-Whitney $p = .553$ and $p = .977$,
respectively, Holm-corrected). A supplementary TOST equivalence analysis ($\Delta = \pm 0.5$; session duration $p_{TOST} = .049$, time-to-peak $p_{TOST} = .026$). This demonstrates that the added interface complexity of the three-slider condition neither accelerated the onset of severe symptoms (time-to-peak) nor reduced participants' overall tolerance to continue the simulation (session duration). Participants in the
three-slider condition made substantially more slider adjustments overall
and per minute than single-slider participants ($p < .001$
Holm-corrected). While this is expected given
that they operated three controls rather than one, it serves as a useful
engagement check confirming that participants actively used all three
sliders rather than defaulting to a single one.
 
\begin{table}[h]
  \caption{Slider usage patterns by condition (Mann-Whitney $U$,
  Holm-corrected; $r$ = rank-biserial effect size).}
  \label{tab:usage}
  \small
  \begin{tabular}{lrrrr}
    \toprule
    \textbf{Metric} & \textbf{1-slider (Mdn)} & \textbf{3-slider (Mdn)} & \textbf{$p$ (Holm)} & \textbf{$r$} \\
    \midrule
    Session duration (min)     & 18.62 & 19.68 & .553    & .10 \\
    Time to reach peak (min)   & 12.71 & 12.16 & .977    & .03 \\
    Total slider adjustments   & 440   & 1517  & $<.001$ & .68 \\
    Adjustments / minute       & 26.0  & 93.3  & $<.001$ & .68 \\
    \bottomrule
  \end{tabular}
  \Description{A table comparing session duration, time to reach peak sickness, and slider adjustment counts between the single-slider and three-slider conditions, with Mann-Whitney p-values and rank-biserial effect sizes. Duration and time-to-peak show negligible effect sizes, while adjustment counts show large effects.}
\end{table}
 
\subsection{Demographic Predictors of Symptom Severity}
To examine whether demographic factors predicted cybersickness severity,
we correlated each participant's continuous slider data (peak severity and area under the curve) with six pre-session variables: age, gender, prior
VR experience, gaming experience, corrective lens use, and prior motion
sickness susceptibility. We used Mann-Whitney $U$ tests for two-group
comparisons. For peak severity, we used the peak of the Oculomotor slider in the single-slider condition. In the three-slider condition, peak severity was defined as the highest peak across the Nausea, Disorientation, and Oculomotor sliders.

\begin{figure}[h]
  \centering
  \includegraphics[width=\linewidth]{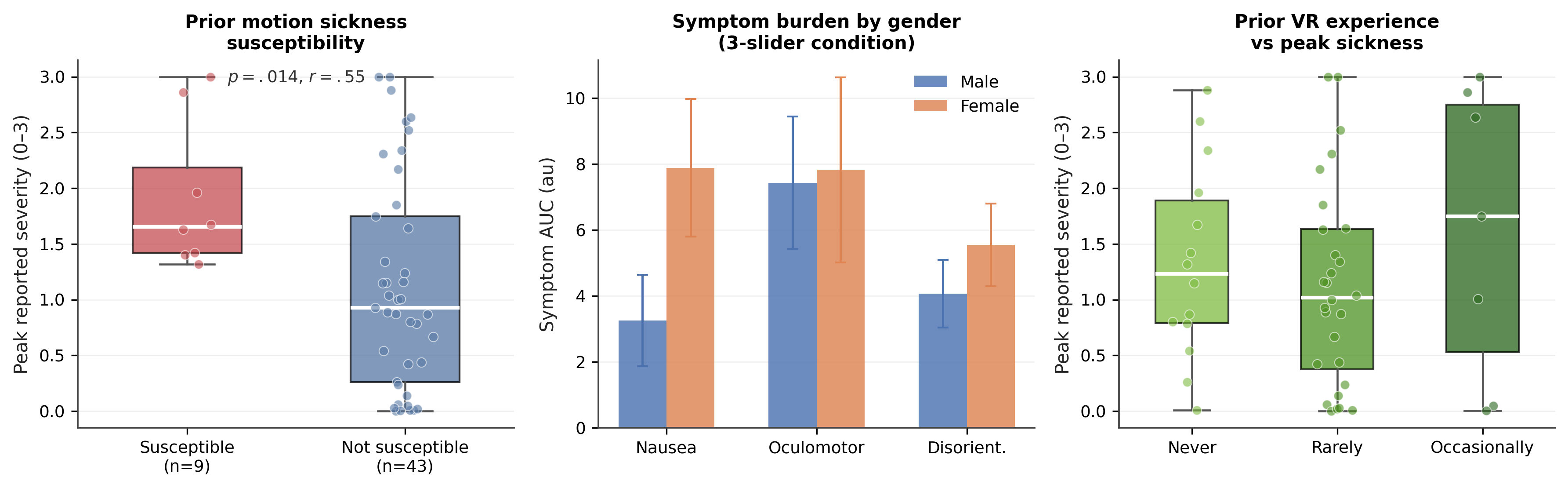}
  \caption{Individual differences in symptom reporting. Left: peak
  sickness by prior motion sickness susceptibility. Middle: symptom AUC
  by gender across the three subscales (3-slider condition). Right: peak
  sickness by prior VR experience. Boxes show median and IQR; dots show
  individual participants.}
  \label{fig:demosick}
  \Description{Three side-by-side charts. The left panel shows clearly higher peak sickness in motion-sickness-susceptible participants. The middle panel shows grouped bars for nausea, oculomotor, and disorientation AUC by gender with overlapping error bars. The right panel shows box plots for three levels of prior VR experience with overlapping distributions.}
\end{figure}

Prior motion sickness susceptibility was the only significant
predictor. Participants who reported a history of motion sickness
reached substantially higher peak severity ($M = 1.91$, $SD = 0.66$) than
those who did not ( $M = 1.12$, $SD = 0.95$), a large and
statistically reliable difference ($U = 255$, $p = .014$; Holm corrected, $r = .55$;
Figure~\ref{fig:demosick}). This indicates that individuals generally
prone to motion sickness also experience stronger cybersickness in VR,
and the slider interface was sensitive enough to capture this at
the subscale level. All other demographic variables like age, gender, prior VR experience,
gaming experience, and corrective lens use showed no significant
association with sickness severity. Female participants
trended higher on nausea AUC in the three-slider condition ($r = .33$)
but this did not reach significance.

We also verified that the two interface conditions were demographically comparable, since any imbalance could confound the between-subjects comparison. After Holm correction for multiple comparisons, no demographic feature differed significantly between the single- and three-slider groups: age (Mann--Whitney $p = .096$, $r = .28$), gender (chi-square $p = .382$), prior VR experience (chi-square $p = .399$), gaming experience (Mann--Whitney $p = .983$, $r = .01$), corrective lens use (Fisher's exact $p = .088$), and prior motion sickness susceptibility (Fisher's exact $p = .138$). These results suggest that random assignment produced well-matched groups with respect to the measured demographic and experiential characteristics. We used Mann--Whitney $U$ tests for two-group comparisons of continuous variables, chi-square tests for categorical variables, and Fisher's exact test for binary variables where small cell counts made the chi-square approximation unreliable \cite{kim2017statistical}. $p$-values were adjusted using the Holm procedure to account for multiple comparisons.

\subsection{Correlation with SSQ (H1, H2)}
 \begin{figure}[htbp]
  \centering
  \includegraphics[width=\linewidth]{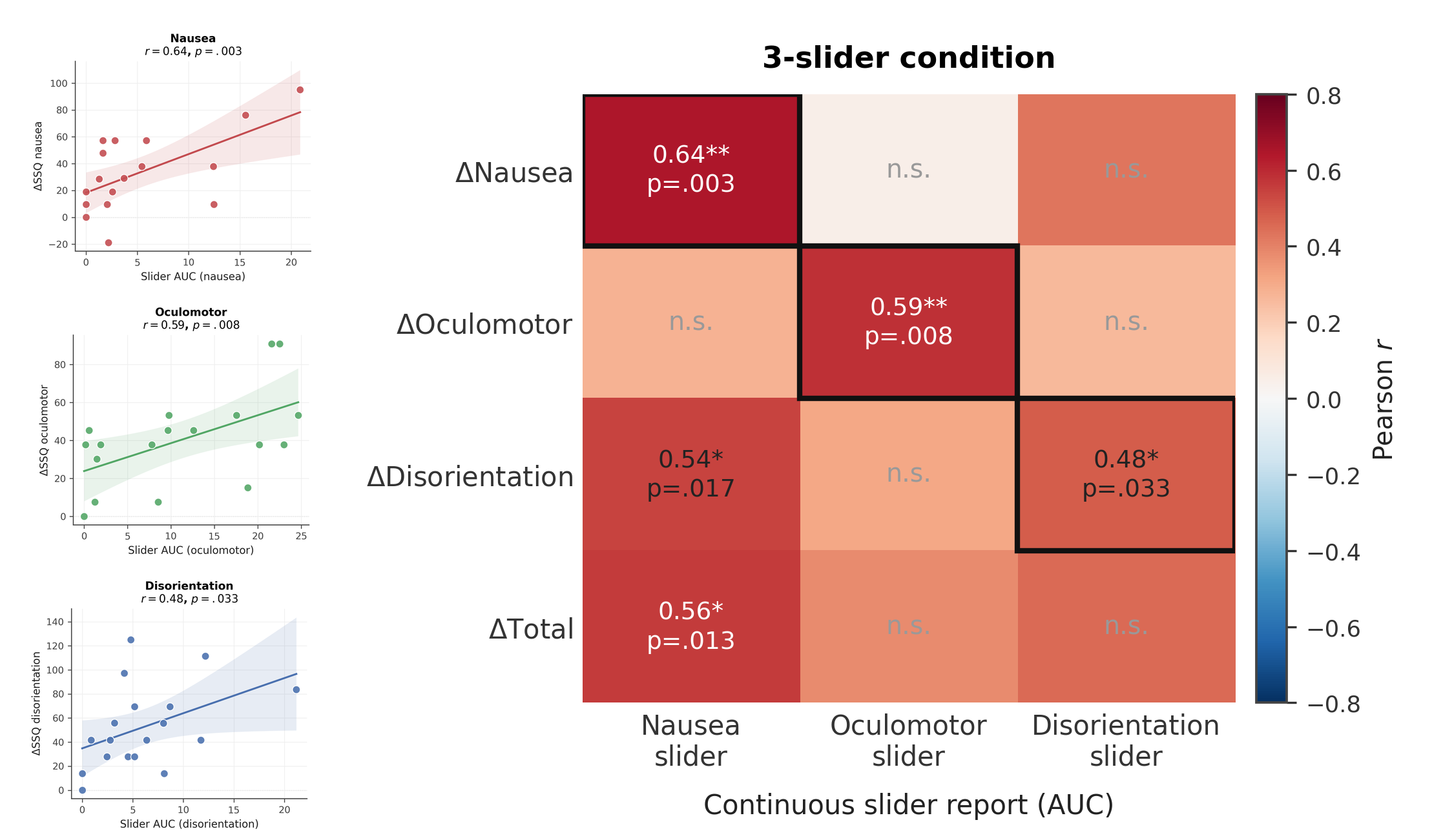}
\caption{Correlation between continuously reported slider values and SSQ subscale changes in the three-slider condition. Left: Scatter plots showing the relationship between each subscale's slider area-under-curve (AUC) and its corresponding $\Delta$SSQ subscale score. Right: Correlation matrix between all three sliders AUC and all $\Delta$SSQ subscales (nausea, oculomotor, disorientation) and total score, with matching diagonal pairs outlined. (supports H1). Asterisks indicate significance: *$p<.05$, **$p<.01$ (Holm corrected).}
  \label{fig:ssqscatter}

\end{figure}
To test H1, we computed Pearson correlations, for participants in the three-slider condition, between each continuously reported subscale (nausea, oculomotor, disorientation) and its corresponding SSQ subscale score, computed as the change from pre- to post-exposure ($\Delta$SSQ subscale). For each participant, we used the area under curve (AUC) of continuously reported value for each subscale across the full duration of the simulation as the summary statistic of interest \cite{titensor2025estimating}. As we can see in the Figure \ref{fig:ssqscatter}, Each slider correlated most strongly with its own matched subscale: nausea ($r = .64$, 95\% CI [.34, .82], $p = .003$), oculomotor ($r = .59$, 95\% CI [.26, .80], $p = .008$, Holm corrected), and disorientation ($r = .48$, 95\% CI [.11, .73], $p = .03$; Holm corrected). Separation was weaker between nausea and disorientation --- the nausea slider also had a strong correlation with $\Delta$disorientation ($r = .54$, 95\% CI [.19, .77], $p = .017$, Holm corrected). Three scatter plots on the left show slider area-under-curve against change in SSQ score for the nausea, oculomotor, and disorientation subscales. Each shows a positive linear relationship with a fitted regression line and shaded confidence band.
\begin{figure}[h!]
\centering
\includegraphics[width=\linewidth]{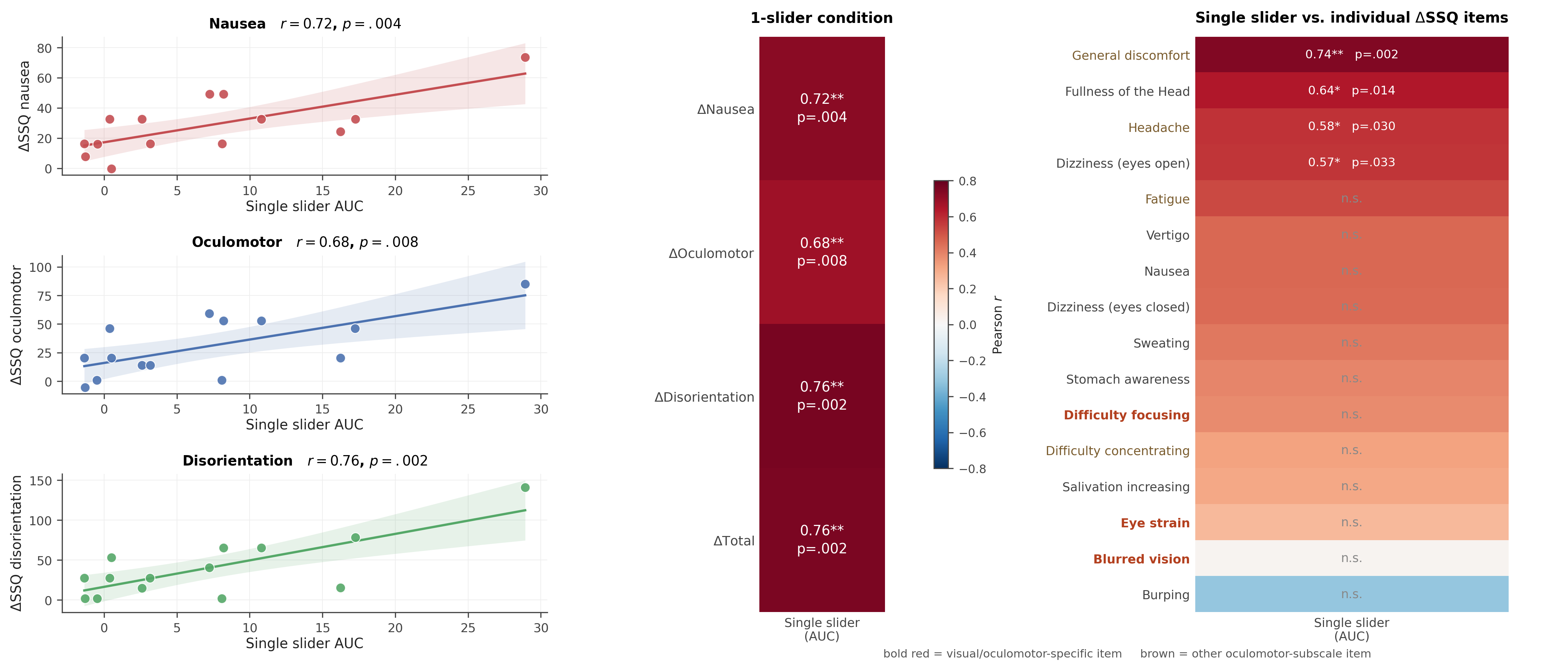}
\caption{Correlation between the continuously reported oculomotor slider value and SSQ subscale changes in the single-slider condition. Left: Scatter plots showing the relationship between the slider area-under-curve (AUC) and $\Delta$Nausea, $\Delta$Oculomotor, $\Delta$Disorientation, and $\Delta$Total SSQ scores. Middle: Correlation matrix with Slider AUC and $\Delta$ SSQ subscales and Right: Correlation of single slider AUC with individual SSQ items. Asterisks indicate significance: *$p<.05$, **$p<.01$. (Holm corrected)}
\label{fig:ssq_slider_corr_1slider}
\label{fig:1 slider}
\end{figure}
According to H2, the single slider condition should be strongly correlated with the oculomotor since we told the participants to change that slider based on the oculomotor symptoms only. However, the slider correlated strongly with each subscale ($r = .68$--$.76$, all $p < .01$; Holm corrected; Figure~\ref{fig:1 slider}). Breaking the SSQ into its 16 individual symptoms explains why (Figure~\ref{fig:1 slider}): the slider was closely correlated with general discomfort- which is an overlapped symptoms of Nausea and Oculomotor. While headache, an oculomotor symptom, also correlated with the slider, distinctly oculomotor items --- difficulty focusing, eye strain, and blurred vision --- did not reach significance. The single slider therefore captured general discomfort rather than oculomotor discomfort specifically, and we reject H2.

\begin{figure}[h]
  \centering
  \includegraphics[width=0.52\linewidth]{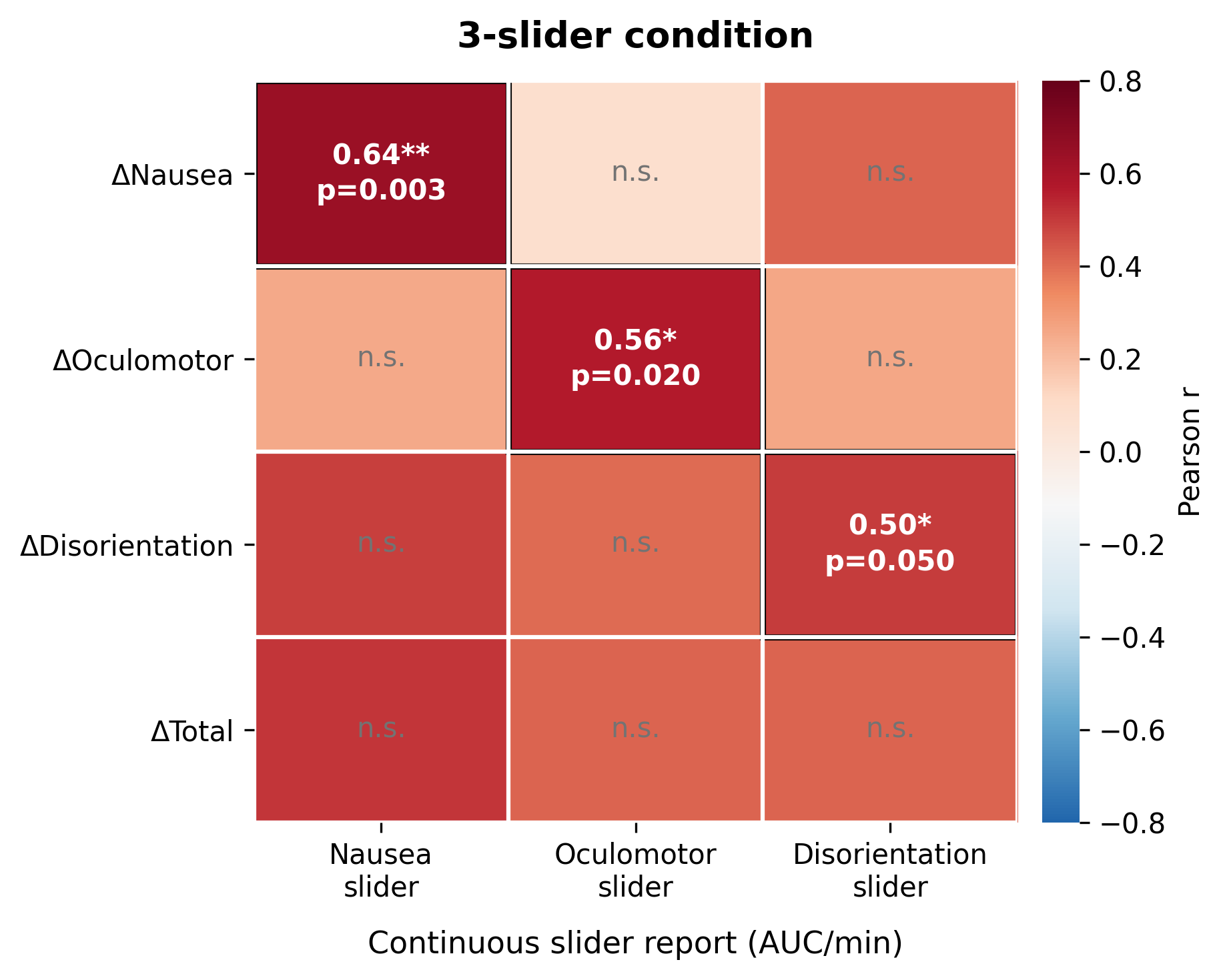}
  \hfill
  \includegraphics[width=0.33\linewidth]{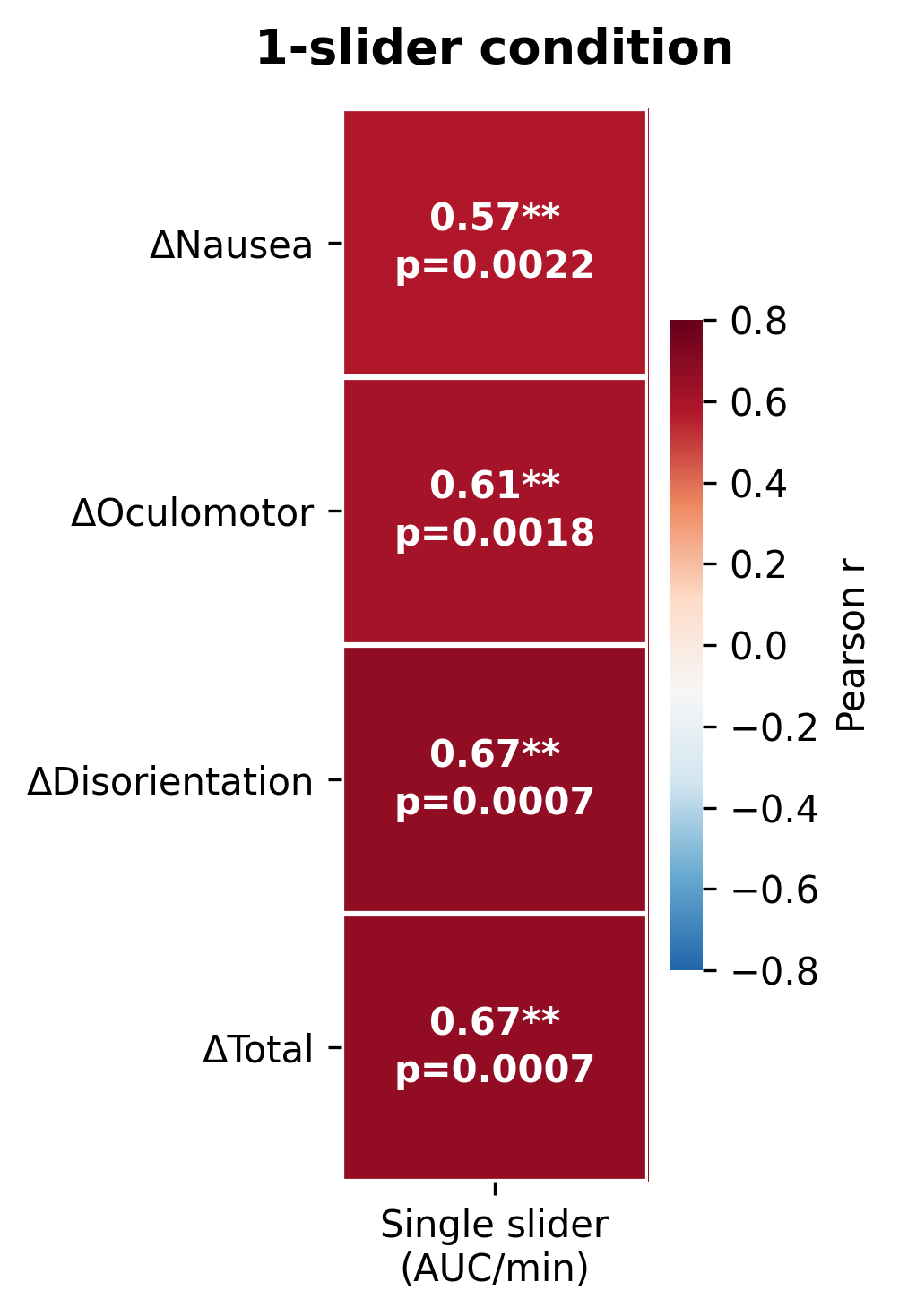}
  \caption{Correlations between exposure-normalized slider values (AUC per minute)
  and $\Delta$SSQ subscale changes. \textbf{Left:} three-slider condition, with
  matched slider--subscale pairs outlined. \textbf{Right:} single-slider condition.
  Asterisks indicate significance after Holm correction ($^{*}p \leq .05$,
  $^{**}p < .01$). }
  \label{fig:aucmin}
\end{figure}

Since total AUC may still be partly influenced by session duration, which could affect the observed correlation coefficients, we repeated all H1 and H2 correlation analyses using AUC normalized by session duration (AUC per minute) as shown in the Figure \ref{fig:aucmin}.
In the three-slider condition, each slider remained most strongly associated with
its matched $\Delta$SSQ subscale (nausea $r = .64$, oculomotor $r = .56$,
disorientation $r = .50$). In the single-slider condition, the slider again correlated
significantly with all four $\Delta$SSQ measures ($r = .57$--$.67$), most strongly
with $\Delta$Disorientation rather than the instructed $\Delta$Oculomotor subscale.
 After accounting for session duration, no off-diagonal pair were significan. This makes the separation more clear and further supports H1. However, H2 remains unchanged: correlating significantly with all the subscales of SSQ.

\subsection{Cognitive Load (H3)}

To test H3, we compared secondary-task (Simon task) performance between the single-slider and three-slider conditions. For each participant, we computed mean reaction time (RT) and accuracy, separately for congruent and incongruent trials. 
\begin{figure}[h]
  \centering
  \includegraphics[width=0.6\linewidth]{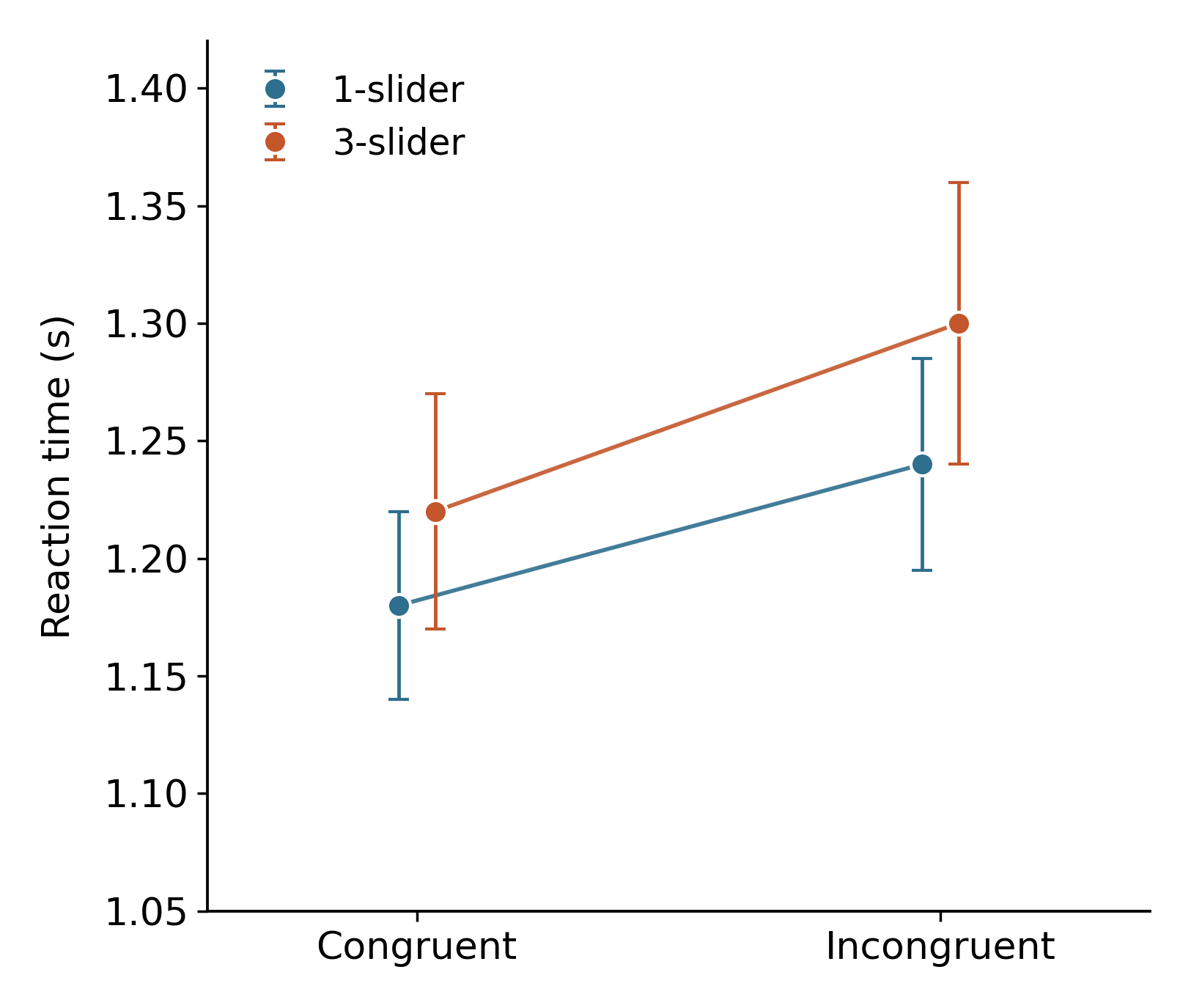}
  \caption{Reaction time by interface and trial type ($\pm$ SE). The
  one-slider and three-slider conditions track each other closely.}
  \label{fig:groupcong}
  \Description{Grouped bars of reaction time for congruent and incongruent trials in each interface condition; the one-slider and three-slider bars are nearly identical, both showing the congruency slowdown.}
\end{figure}
Splitting reaction time by interface (Figure~\ref{fig:groupcong}) showed the one-slider and three-slider conditions tracking each other closely at both trial types. Comparing the two conditions
on each Simon measure with Mann-Whitney $U$ tests revealed no significant
differences, with small effect sizes throughout (reported as rank-biserial
$r$): accuracy ($p = .97$, $r = .19$), congruent reaction time
($p = 1.00$, $r = .08$), incongruent reaction time ($p = 1.00$, $r = .09$),
and the congruency effect ($p = 1.00$, $r = .09$; Holm-corrected;
Figure~\ref{fig:groupcong}). To test
this formally while separating the effect of congruency from the effect of
the interface, we fit an ordinary least-squares (OLS) regression predicting
reaction time from congruency, interface group, and their interaction.
Because each participant contributed two data points (one per trial type),
these observations are not independent, so we used standard errors clustered
by participant~\cite{cameron2015robust}: 
\begin{equation}
\mathrm{RT} = \beta_0
            + \beta_1(\text{Congruency})
            + \beta_2(\text{Group})
            + \beta_3(\text{Congruency} \times \text{Group}),
\end{equation}
where \textit{Congruency} is 0 for congruent and 1 for incongruent trials,
and \textit{Group} is 0 for the 1-slider and 1 for the 3-slider interface.

\begin{itemize}
\item $\beta_1$ (\textit{Congruency}) captures the Simon interference
effect: reaction time was significantly longer on incongruent than
congruent trials ($\beta_1 \approx 57$,ms, 95\% CI [0.030, 0.083], $p < .001$), confirming the
task engaged attention as intended.

\item $\beta_2$ (\textit{Group}) captures the overall difference between
interfaces: participants using the 3-slider interface were about 35,ms
slower on average, but this difference was not significant ($\beta_2 = 0.0347$, 95\% CI [-0.096, 0.166], $p = .60$).

\item $\beta_3$ (\textit{Congruency} $\times$ \textit{Group}) tests whether the congruency slowdown is larger
under the 3-slider interface, which would indicate the extra sliders
consume attention. The slowdown was about 25,ms larger in the 3-slider
group, but this interaction was not significant ($\beta_3 = 0.0248$, 95\% CI [-0.034, 0.083], $p = .41$).
\end{itemize}
To determine whether these small differences constituted practical
equivalence, we conducted a TOST analysis using an equivalence bound of
$d = \pm0.5$. We defined equivalence bounds of $d = \pm 0.50$, corresponding to Cohen's conventional benchmark for a medium standardized mean difference~\cite{cohen2013statistical}, as our equivalence bound. All four TOST $p$-values failed to reach significance ($p$
ranging from $.106$ to $.227$), yielding an inconclusive verdict.
Taken together, the uniformly small observed effect sizes (rank biserial r) provide no evidence that the three-slider interface increased attentional or cognitive load. However, the equivalence test was inconclusive at the conventional equivalence bound, indicating that a larger sample is needed to establish true equivalence between the two interfaces. Therefore, we consider H3 partially supported: while the three-slider interface did not show evidence of increased cognitive load, we cannot conclusively claim that the two interfaces are equivalent.
 
\subsection{Relationship Between Cybersickness and Attention (H4)}
\begin{figure}[h!]
    \centering
    \begin{minipage}{0.62\textwidth}
        \centering
        \includegraphics[width=\linewidth]{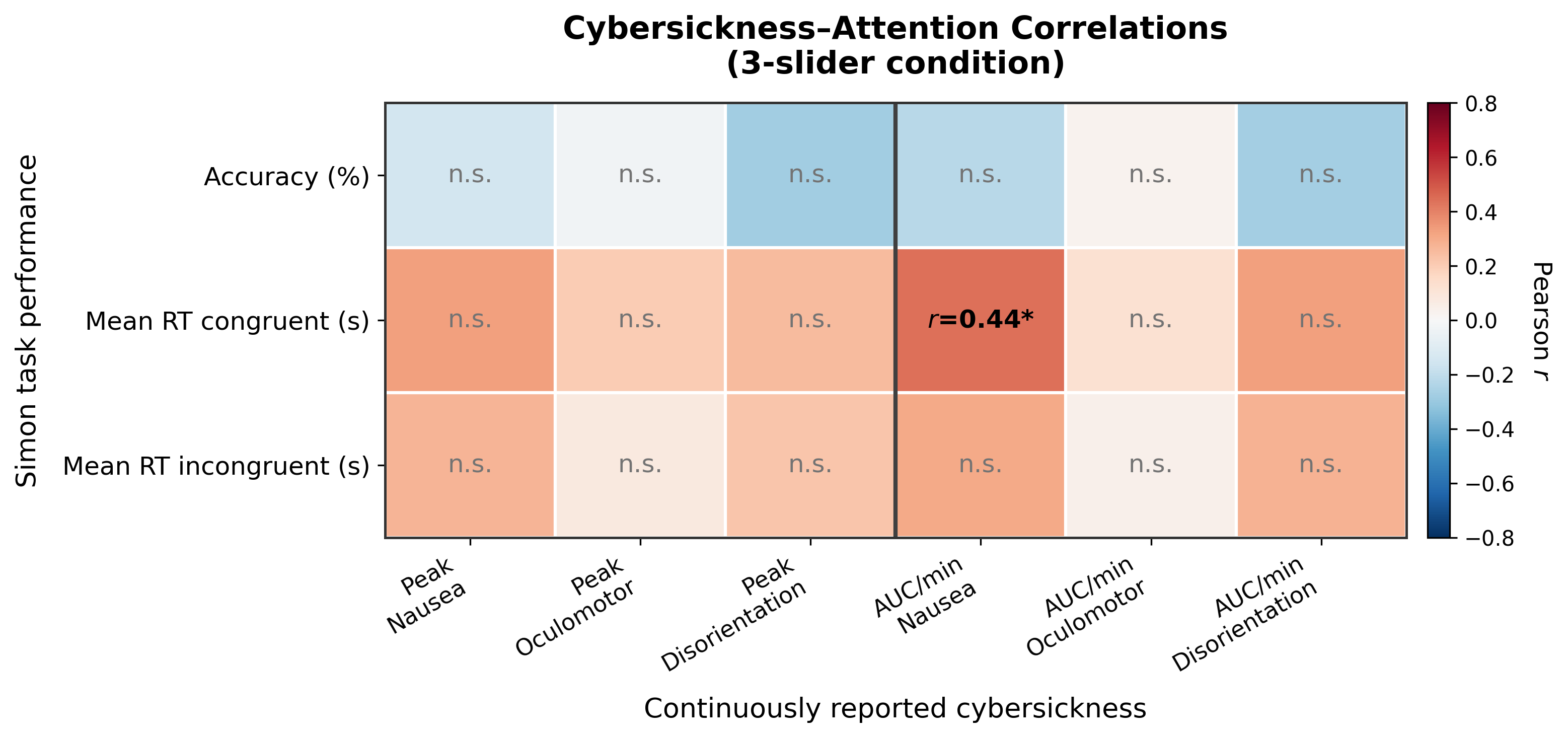}
    \end{minipage}\hfill
    \begin{minipage}{0.37\textwidth}
        \centering
        \includegraphics[width=\linewidth]{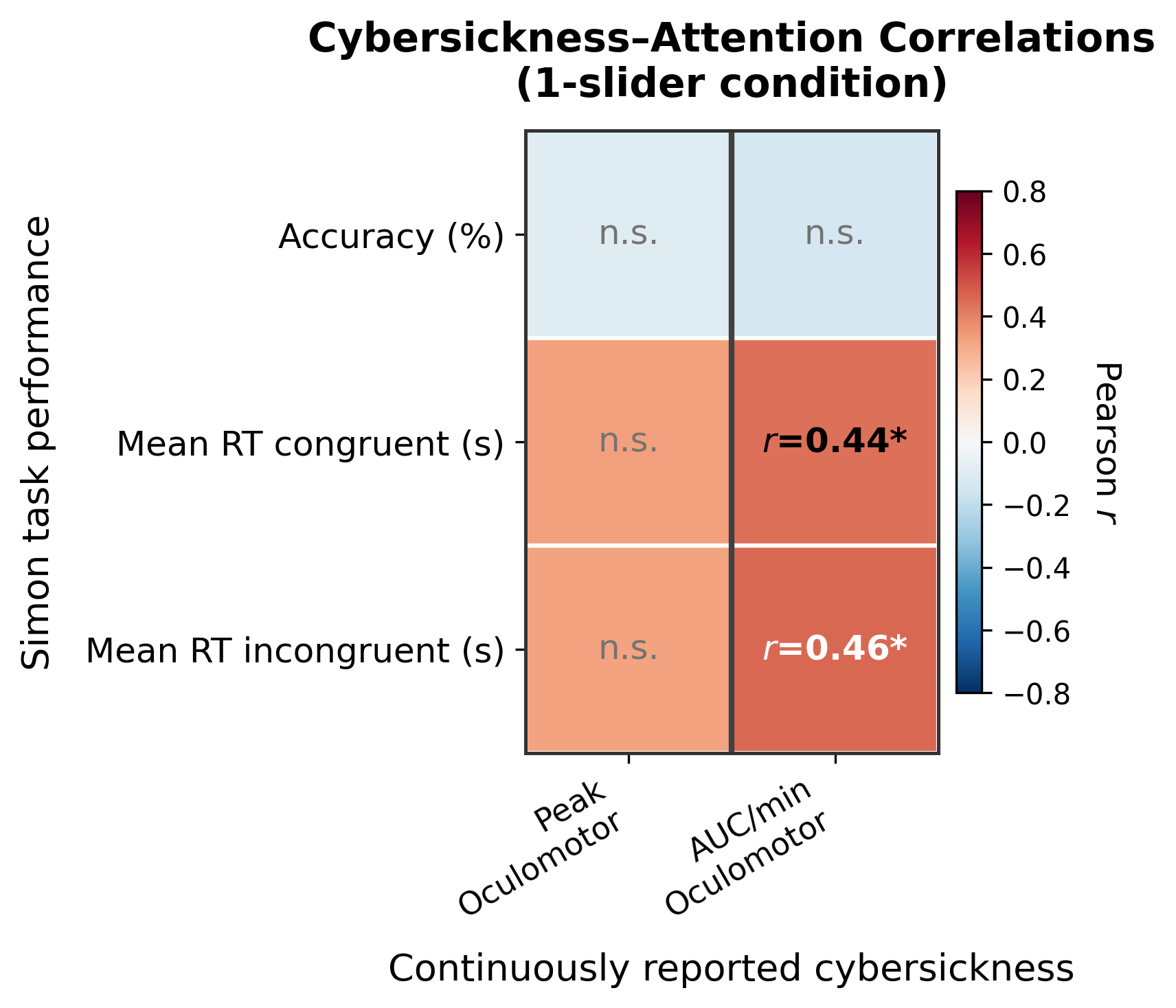}
    \end{minipage}
    
    \caption{Correlation between continuously reported cybersickness and Simon task performance. \textbf{Left:} Three-slider condition ($N=26$). \textbf{Right:} Single-slider condition ($N=26$). *$p<.05$.}
    \label{fig:h4_heatmaps}
\end{figure}
To test H4, we examined the relationship between reported cybersickness severity
and concurrent Simon task performance. Because the single-slider interface
collected only oculomotor ratings, whereas the three-slider interface collected
nausea, oculomotor, and disorientation ratings, we analyzed the two conditions
separately. For each participant we computed summary cybersickness metrics
(area under the curve per minute, AUC/min, for each available subscale) and
summary Simon performance (accuracy, mean RT for congruent trials, and mean RT
for incongruent trials) across the full session, and correlated these across
participants. This is a between-person analysis: it tests whether participants
who reported more sickness performed worse overall.

In the three-slider condition, nausea AUC/min was positively associated with mean RT on congruent trials ($r = .44$, 95\% CI $[.11, .70]$, $p = .026$, Holm-corrected, $N = 26$; Figure~\ref{fig:h4_heatmaps}). Thus, participants who experienced greater cumulative nausea tended to respond more slowly on congruent Simon trials. RT was also positively correlated with the other cybersickness subscales as expected, while accuracy was generally negatively correlated with sickness severity; however, none of these additional correlation reached statistical significance.

In the single-slider condition, oculomotor AUC/min was positively associated with mean RT for both congruent trials ($r = .44$,  95\% CI $[.11, .70]$, $p = .029$, Holm-corrected) and incongruent trials ($r = .46$, 95\% CI $[.14, .72]$, $p = .020$, Holm-corrected; $N = 26$; Figure~\ref{fig:h4_heatmaps}). Peak oculomotor severity showed similar positive, but non-significant, associations with RT for congruent ($r = .33$) and incongruent trials ($r = .32$). Neither peak oculomotor severity nor oculomotor AUC/min was significantly associated with accuracy.

This correlation analysis tests whether participants who reported more sickness performed worse overall. To consider the moment-to-moment detail, we also analysed performance at the level of individual Simon trials. We therefore fit a linear
mixed-effects model \cite{west2022linear}, with trials nested within participants. For each trial, we assigned the average slider sickness rating from the 30-second window in which the trial occurred, and examined whether sickness severity predicted reaction time. The model included a random intercept for each participant to account for differences in
participants' baseline performance, and trial congruency as a fixed effect to control for the usual Simon interference effect. For trial $i$ from participant $j$:
 
\begin{equation}
\mathrm{RT}_{ij} = \beta_0 + \beta_1 \mathrm{Severity}_{ij}
  + \beta_2 \mathrm{Congruent}_{ij} + u_j + \varepsilon_{ij},
  \qquad u_j \sim \mathcal{N}(0, \tau^2)
\end{equation}
 
\noindent
where $\mathrm{Severity}_{ij}$ is the window-mean slider rating (0.0--3.0) and
$\mathrm{Congruent}_{ij}$ is 1 for congruent and 0 for incongruent trials. The term
$u_j$ is the participant-specific random intercept, which captures stable differences
in each participant's baseline response speed, and $\varepsilon_{ij}$ is the
trial-level residual. 

Accuracy was analysed in the same way using a binomial mixed model with a logit link. This
approach kept all approximately 5,200--5,300 trials in each panel as separate
observations, rather than averaging them across participants. We also included a
severity $\times$ congruency interaction to test whether sickness severity had a
different effect on performance during congruent versus incongruent trials.

\begin{figure}[h!]
  \centering
  \includegraphics[width=\linewidth]{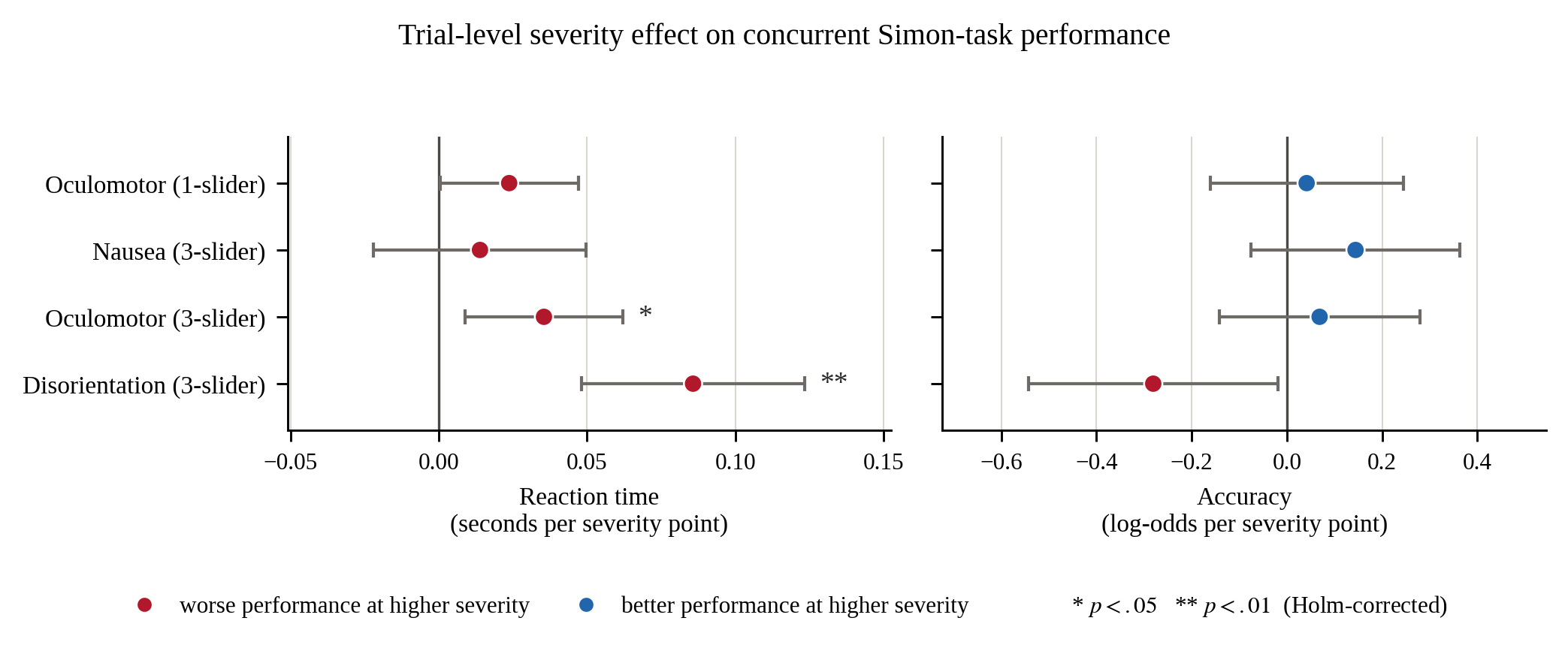}
  \caption{Fixed-effect estimates from the trial-level mixed models relating
  window-mean cybersickness severity to concurrent Simon-task performance (random
  intercept per participant, congruency controlled; $\approx$5,200--5,300 trials from
  26 participants per panel). Points show the change in performance per one-point
  increase in reported severity; bars are 95\% confidence intervals. Red indicates an
  effect in the direction of impaired performance. $^{*}p < .05$, $^{**}p < .01$}
  \Description{Forest plot of severity coefficients for reaction time and accuracy
  across the four condition-subscale panels.}
  \label{fig:severity-forest}
\end{figure}

The coefficient $\beta_1$ is the change in reaction time per one-point rise on the 0--3 slider, in seconds, where positive values indicate slower responses. All four
setup showed slower responses at higher reported severity
(Figure~\ref{fig:severity-forest}). The strongest relation was found for disorientation in the three-slider condition ($\beta_1 = 0.086$\,s per severity
point, 95\% CI [0.048, 0.124], $p < .001$, Holm-corrected), followed by
oculomotor discomfort in the three-slider condition ($\beta_1 = 0.035$\,s, 95\%
CI [0.009, 0.062], $p = .009$, Holm-corrected). Single-slider oculomotor
($\beta_1 = 0.024$\,s, $p = .09$, Holm-corrected) and three-slider nausea
($\beta_1 = 0.014$\,s, $p = .448$, Holm-corrected) were not significant. For
accuracy, no subscale was significantly associated with severity after Holm
correction. The severity $\times$ congruency interaction was insignificant for all the slider setup indicating that severity didn't affect congruent and incongruent trials differently.

Taken together, these findings partially support H4. We hypothesized that greater
symptom severity would be associated with both slower reaction times and lower
accuracy. The reaction-time component was
supported but no accuracy association survived correction in either
analysis. We
therefore conclude that reported cybersickness was reliably associated with slower responding, but not with reduced accuracy, on the concurrent Simon task.

\subsection{Usability (H5)}

To test H5, we compared the ease of use of the two reporting interfaces.
Participants took two standard self-report questionnaires administered
immediately after the session: SUS and the
NASA-TLX. SUS consists of ten items rated on a 1--5
scale, alternating between positive (e.g., ``I found the system easy to
use'') and negative (e.g., ``I found this system unnecessarily complex'')
statements \cite{lima2021severity}. Responses are converted to a composite score from 0 to 100,
where scores above 68 are generally considered acceptable. We computed a
SUS score for each participant and compared the two conditions using a
 Mann-Whitney $U$ test.
Both conditions were rated as usable and comfortable to operate. The
three-slider group scored $M = 74.2$ ($SD = 14.9$, Mdn $= 77.5$) and the
single-slider group scored $M = 73.4$ ($SD = 16.2$, Mdn $= 75.0$), both
well above the acceptability threshold. The difference was negligible
($U = 560$, $p = 1.00$, $r = .00$).
\begin{figure}[h]
  \centering
  \includegraphics[width=\linewidth]{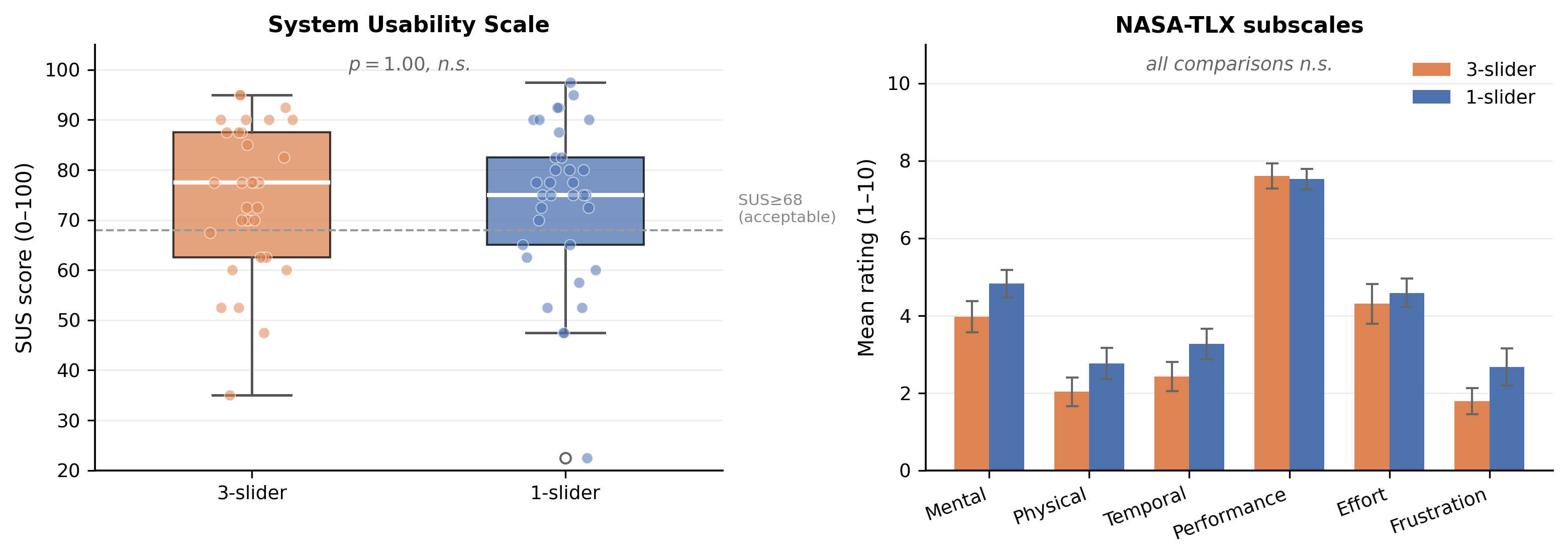}
  \caption{Usability outcomes by condition. Left: SUS scores (boxes show
  median and IQR; dots show individual participants; dashed line marks the
  SUS acceptability threshold of 68). Right: NASA-TLX subscale means
  ($\pm$ SE). No comparison reached significance, and the two interfaces
  were statistically equivalent on the SUS.}
  \label{fig:usability}

\end{figure}
Because a non-significant test does not by itself demonstrate that the two
interfaces are equally usable, we followed up with a TOST equivalence analysis, using an equivalence bound of Cohen's
$d = 0.5$~\cite{lakens2018tost}. The SUS scores were statistically
equivalent ($p = .035$): the observed difference ($d = 0.05$) fell well
within the equivalence bounds, allowing us to positively conclude that the
two interfaces are equally usable, directly supporting H5.

\textit{NASA-TLX} measures six dimensions of perceived workload --- mental
demand, physical demand, temporal demand, performance, effort, and
frustration --- each rated on a 1--10 scale \cite{rubio2004evaluation, hoonakker2011measuring}. We compared each subscale
between conditions using Mann-Whitney $U$ tests (Figure~\ref{fig:usability}).
No subscale differed significantly (all $p > .10$, all $r \leq .23$). The three-slider condition trended toward \textit{lower} workload
on mental demand, temporal demand, and frustration, though these differences
did not reach significance.
A follow-up TOST at the $d = 0.5$ bound found the two conditions
statistically equivalent on perceived performance ($p = .034$) but was
inconclusive for the remaining subscales, so we do not claim equivalence for
those. Together, these results partially support H5: the two interfaces were comparable in
usability, showing statistically equivalent SUS scores and equivalent
perceived performance on the NASA-TLX, with no evidence that the
three-slider interface increased workload on any dimension.

\subsection{Correlation with Different Modalities}
To support an exploratory analysis of modality--symptom relationships, we extracted a common set of windowed features from all five recorded data streams: eye tracking, head tracking, ECG, PPG, and GSR. Each stream was first cleaned and filtered independently, then summarized within a sliding 30~s window. A detailed description of the signal processing and feature extraction pipeline for each modality is provided in Appendix~\ref{sec:appendix-preprocessing}. Figure~\ref{fig:sample-window} shows a representative 30~s window of processed multimodal data from a single participant in Appendix~\ref{app:fig}.

Because each participant contributed many windows, correlating pooled windows
would conflate between-participant differences with the within-person
relationship of interest, and would treat non-independent observations as
independent, inflating significance. We therefore used repeated-measures
correlation (rmcorr)~\cite{bakdash2017rmcorr}, which estimates a single common
slope across participants while allowing each participant their own intercept.
\begin{figure*}[h]
  \centering
  \includegraphics[width=\linewidth]{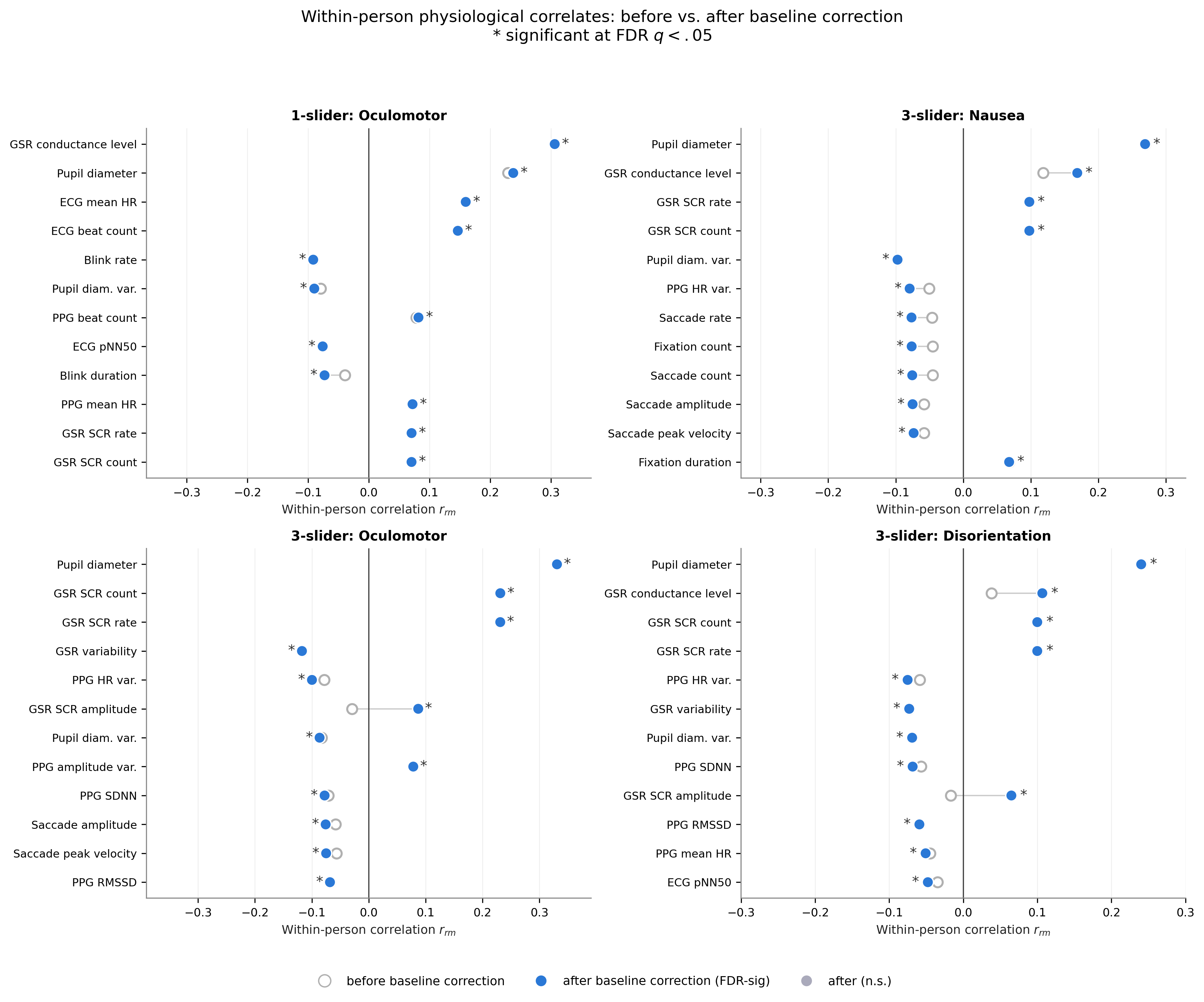}
  \caption{Within-person (rmcorr) correlations between physiological features and reported symptom severity, shown before and after baseline correction, for each condition and subscale. Hollow points show the correlation 
 before baseline correction and filled points after; a connecting line marks how much the correction shifted each estimate (for most features the two coincide). (*) marks means significant at FDR correction}
  \label{fig:featcorr}

\end{figure*}
Before correlation, each physiological feature was expressed relative to the
participant's pre-exposure VR baseline. VR baseline was used because we wanted
to take into account how much it changes from non-stimulating VR environment
since we are considering cybersickness. As a preliminary check, we also tested
whether any baseline feature differed between the two interface conditions with
a Mann--Whitney $U$ test. After Holm correction, no feature showed a
significant difference (all $p > .05$), indicating that the two groups did not
differ before the simulation started.

For every physiological and eye-tracking feature, we computed a
participant-specific baseline mean and standard deviation using their two-minute
VR baseline \cite{seropian2022comparing, mathot2018safe}. Each
simulation-window feature was then converted into a within-participant
$z$-score against this baseline. Head-tracking features were excluded, as no
VR-baseline equivalent exists to compute a reference.

We report the within-person correlation $r_{rm}$ for each feature--subscale
pair, correcting $p$-values for multiple comparisons with the
Benjamini--Hochberg false discovery rate (FDR) within each subscale and
treating $q < .05$ as significant \cite{BenjaminiHochberg1995}. In the
single-slider condition, each feature was related to the oculomotor rating; in
the three-slider condition, each feature was related separately to the nausea,
oculomotor, and disorientation ratings
(Figure~\ref{fig:featcorr}). 

Pupil diameter showed the strongest and most stable
association: it correlated positively with all four symptom measures
(single-slider oculomotor $r_{rm} = .24$; three-slider nausea $.27$,
oculomotor $.33$, disorientation $.24$; all $q < .05$), indicating that
participants' pupils dilated as their reported sickness rose.
Skin conductance level and spontaneous skin-conductance responses (SCR rate and count) increased with reported symptoms across conditions (e.g., SCL $r_{rm} = .31$ for single-slider oculomotor; SCR count and rate $r_{rm} = .23$ for three-slider oculomotor; all $q < .05$), further supporting the use of physiological measures as indicators of cybersickness severity \cite{brown2026physiological}.

\section{Discussion}
 \subsection{Dataset novelty}
\label{sec:disc-dataset}
In this study, we introduced a custom three-slider physical interface designed to capture continuous, symptom-specific cybersickness dimensions (nausea, oculomotor discomfort, and disorientation) during VR exposure. While recent research has shifted toward continuous rather than discrete reporting—demonstrating that continuous methods can capture sickness fluctuations with greater precision \cite{cantory2025feasibility}. Most of these methods have relied on a single scale to measure overall discomfort, such as the FMS scale \cite{cantory2025feasibility, ryu2026predicting}. Plaza et al. recently evaluated a three-scale approach (nausea, disorientation, and oculomotor discomfort) similar to ours. However, their work primarily focused on comparing different virtual UIs  and they use a single joystick rather than independent physical controls.\cite{plaza2026design}. In contrast, our study specifically validates whether continuous reporting across three physical sliders is consistent with established SSQ subscales, and explicitly investigates whether managing three sliders instead of one adversely affects user attention and cognitive load.

Public cybersickness datasets typically provide a single overall sickness measure. To our knowledge, ours is the first to release independent continuous ratings of nausea, oculomotor discomfort, and disorientation, together with eye tracking, head tracking, GSR, ECG, PPG, and concurrent Simon-task reaction time and accuracy. This combination enables research questions that are difficult to examine using existing datasets, such as identifying which physiological signals are most strongly associated with specific cybersickness symptoms rather than cybersickness treated as a single overall construct.
 The dataset also provides two additional features that may support future research. First, our sample covers a relatively broad age range (18--49 years, $M = 28.3$, $SD = 7.9$), although it remains skewed toward a student demographic since the majority of participants were recruited from a university population. We also examine demographic and individual-difference factors, including age, gender, prior VR and gaming experience, corrective lens use, and previous motion-sickness history, and report their relationships with cybersickness severity. This information can support research into individual differences in cybersickness susceptibility rather than focusing only on physiological manifestations of sickness.
Second, we provide synchronized baseline recordings collected before immersion for every modality. These include both an outside-VR seated baseline and a VR-baseline. These recordings allow future researchers to examine within-participant changes and physiological reactivity relative to each individual's baseline, rather than relying solely on absolute measurements or comparisons across participants. None of the three prior datasets listed in provides both of these baseline conditions, making them an additional feature of our dataset.

\subsection{Three sliders correlates with corresponding symptoms}

Supporting H1, every slider in the three-slider condition correlated most strongly with its own matched SSQ subscale: nausea ($r = .64$), oculomotor ($r = .59$), and disorientation ($r = .48$), forming a clear diagonal pattern in the correlation matrix (Figure~\ref{fig:ssqscatter}). We used the AUC of each slider trace rather than peak severity, because AUC reflects the accumulated symptom burden across the entire exposure, whereas peak captures only a single moment. This suggests participants could track three symptom dimensions during immersive VR that align with their intended SSQ subscales. Since we considered SSQ as for dividing the subscale and as post session baseline questionnaire to compare, SSQ itself has overlapping symptoms. For example, While calculating Disorientation it also include nausea (which is major symptom of nausea subscale) or difficulty focusing which is also included in oculomotor symptom. Therefore nausea slider also showing a strong correlation with the disorientation SSQ score (r = .54). Part of this overlap appears attributable when we consider exposure duration of participants.  When we divide each slider's AUC by
session length, the link between the nausea slider and $\Delta$Disorientation is no
longer significant. 

Conversely, we reject H2, which hypothesized that the single slider would specifically track oculomotor discomfort. In the single-slider condition, continuous reports correlated strongly with all post-exposure SSQ subscales: Nausea ($r = 0.72$), Oculomotor ($r = 0.68$), and Disorientation ($r = 0.76$). This pattern holds when we account for exposure time: with AUC per minute, the single
slider still correlates significantly with every subscale ($r = .57$--$.67$), and
still most strongly with disorientation rather than oculomotor. An item-level analysis of the 16 individual SSQ symptoms with AUC of slider revealed that the single slider correlated most highly with general discomfort, fullness of the head, headache, and dizziness. While headache is a specifically targeted oculomotor symptom, general discomfort contributes to both the oculomotor and nausea subscales, and fullness of the head and dizziness are primary indicators of disorientation. 
Participants in the single-slider condition were familiarized with oculomotor symptoms and instructed to adjust the slider solely in response to those specific cues. During the post-session open-ended questionnaire, we asked participants whether they experienced symptoms beyond the explained oculomotor ones and if they had adjusted the slider for those uninstructed symptoms. The feedback revealed that a vast majority (86\%) experienced concurrent symptoms, such as nausea and dizziness. Notably, 22\% of participants explicitly stated that they increased the slider in response to these additional symptoms instead of only oculomotor. Thus, these participants effectively used the slider as a measure of overall discomfort rather than solely oculomotor symptoms, which likely contributed to the strong correlations observed across the SSQ subscales.
\subsection{Does a multi-slider interface add usability issues or cognitive load?}
\label{sec:disc-cost}
We hypothesized that attentional demands would be equivalent between the single-slider and three-slider conditions because the Simon task is designed to impose minimal working memory load. Furthermore, the two tasks utilize distinct perceptual modalities and response mechanisms, minimizing cognitive interference: the Simon task relies on auditory input and discrete button presses, whereas slider reporting requires interoceptive monitoring and continuous positional adjustments. Crucially, scaling from one to three sliders merely multiplies an existing activity rather than introducing a qualitatively new attentional demand. Because participants received extensive pre-study training to intuitively map their symptoms to the sliders, tracking three concurrent symptoms did not impose a significantly higher cognitive load than tracking a single symptom \cite{wickens2008multiple}.

The secondary Simon task analysis provides evidence that the three-slider interface did not significantly affect concurrent attention. Reaction times were slower on incongruent than congruent trials in both conditions ($\beta_1 \approx 57$~ms, $p < .001$), confirming that the task produced the expected congruency effect and successfully engaged attention. Overall reaction time did not differ significantly between the single- and three-slider conditions ($\beta_2 \approx 35$~ms, $p = .60$). Accuracy also did not differ significantly between conditions. The potential reason is the simplicity of the Simon task, coupled with pre-study training, allowed participants to maintain high accuracy across all trial types and interface conditions. Although the RT equivalence test was inconclusive, the relatively small Mann--Whitney effect size ( subsection 6.5) suggests that any difference in reaction time between conditions was likely negligible.
Overall, the findings from cognitive load suggest that managing three sliders did not increase perceived difficulty, substantially alter behavior, or impair performance on the concurrent attention task compared with managing a single slider.

Subjective usability (SUS) scores for the three-slider ($M = 74.2$) and single-slider ($M = 73.4$) conditions both exceeded acceptability thresholds and were statistically equivalent, supporting H5. NASA-TLX ratings also showed no significant differences across dimensions, with perceived performance specifically confirmed to be statistically equivalent. 
\subsection{Relation of attention with cybersickness}
\label{sec:disc-attention}
Our dual-task design let us test whether cybersickness was linked to attention (H4). We expected that people who felt sicker would respond more slowly. We tested this in two ways. The correlation analysis asks
whether participants who reported more sickness across the whole session performed worse across the whole session. The trial-level model asks whether performance tracked sickness as it changed during the session.

In the between-person analysis, the one-slider condition showed that greater cybersickness was significantly associated with slower reaction times for both congruent and incongruent trials. However, in the three-slider condition, only nausea severity exhibited a significant correlation with RT in congruent trials; the other specific symptoms were not significantly linked to performance differences. This discrepancy is likely because participants in the one-slider condition treated the single input as a measure of 'overall' cybersickness—a behavior supported by both their questionnaire responses and the data shown in Figure \ref{fig:1 slider}.

The trial-level analysis pointed to different symptoms. Disorientation had the
strongest link with reaction time, followed by oculomotor discomfort in the same
condition. Participants who reported higher nausea overall in the three-slider condition responded more slowly. At the trial level, however, trials occurring in windows with higher reported disorientation had significantly slower responses, whereas the association with nausea was not significant. One explanation is that disorientation is triggered by specific moments of visual–vestibular conflict, such as loops and drops, whereas nausea builds up more gradually over the exposure \cite{reuten2021feelings}. A whole-session summary captures accumulated nausea well, while the 30-second windows are better suited to a symptom that fluctuates. The single-slider condition behaved like nausea in the trial-level analysis. The association with reaction time was positive but not significant, meaning the moment-to-moment changes were not strong enough to detect reliably. Future work could test different window lengths to find the timescale at which each symptom relates most closely to performance. Including time in the session as a predictor would also help separate symptom-driven slowing from general fatigue. Accuracy showed no reliable relation with sickness in either analysis. This is likely because the Simon task was simple and participants were trained beforehand. Sickness slowed congruent and incongruent trials by a similar amount, and the severity × congruency interaction was not significant in any panel. This suggests cybersickness reduces the attention available for the task in general, rather than specifically weakening the ability to suppress a conflicting response.

\subsection{Physiological Correlates of Cybersickness}

Across conditions, pupil diameter and electrodermal activity showed the
most consistent positive associations with cybersickness ratings \cite{brown2026physiological, jeong2022eyes}. Because these
were within-participant correlations, they indicate that changes in a person's
own physiology covaried with changes in their own symptoms over time.

Pupil diameter increased across all three symptom dimensions. This pattern is
consistent with increased autonomic arousal and cognitive demand during
cybersickness \cite{john2019pupil}. Similarly, the positive
association between GSR and oculomotor symptoms likely reflects general
arousal rather than visual strain specifically. The strongest association was between skin conductance level and the
single-slider oculomotor rating ($r_{rm}=.31$). However, our SSQ analysis
suggests that participants used the single slider as a measure of general
discomfort rather than oculomotor strain alone. Thus, this correlation likely
reflects the relationship between overall sickness severity and sympathetic
arousal, for which electrodermal activity is a well-established indicator
\cite{MazloumiGavgani2018Comparative}. The GSR--Nausea association is
more directly compatible with motion-sickness physiology, as increased tonic
skin conductance and phasic electrodermal responses have been reported during
nausea \cite{MazloumiGavgani2018Comparative}.

Baseline normalization most affected electrodermal features. Several GSR
measures changed from weak or negative uncorrected relation to significant
positive correlations after baseline z-scoring. This is plausible because GSR
shows substantial between-person differences in tonic level and may drift over
an experimental session \cite{society2012publication}. Baseline z-scoring
therefore emphasizes each participant's change relative to their own
pre-exposure VR state. Baselines can be established in several ways, including subtracting a resting
value, dividing by it, or standardizing each participant's series relative to
their own baseline mean and variance \cite{Dey2018, mathot2018safe}. We adopted per-participant z-scoring,
which expresses each simulation-window feature in standard-deviation units
relative to that participant's two-minute VR baseline. This reduces both
additive baseline offsets and scale differences between participants; for
example, some individuals show a wider physiological range than others
\cite{mathot2018safe}.
We screened many features against each subscale, we controlled
the false discovery rate with the Benjamini--Hochberg procedure
\cite{BenjaminiHochberg1995}, applied within each subscale. This procedure controls
the expected proportion of false discoveries among results declared
significant, providing a useful balance between false-positive control and
statistical power for exploratory feature screens \cite{BenjaminiHochberg1995}.

\section{Limitations and Future Directions}

Several limitations should be considered when interpreting our findings. First, although our sample size provided sufficient power for the main correlation hypotheses (H1, H2) and the significance tests between conditions, it was underpowered for the equivalence hypotheses (H3, H5) at the conventional equivalence bound of $d = 0.5$ (achieved power = 31\%). This limited our ability to draw strong conclusions about equivalence and may explain why reaction time and NASA TLX did not meet the equivalence criterion. Second, some SSQ symptoms belong to more than one subscale, making it difficult to separate nausea, oculomotor discomfort, and disorientation cleanly. This overlap may have influenced the relationships between the slider ratings and the three SSQ subscales. Third, the single-slider and three-slider conditions differed not only in the number of available controls but also in the symptom information provided during training. Participants in the single-slider condition were trained only on oculomotor symptoms and were instructed to use the slider only for those symptoms, whereas participants in the three-slider condition received training on all three symptom categories, including their potential overlap. Consequently, the rejection of H2 cannot be interpreted as a test of interface dimensionality alone, as the two conditions also differed in instructional content. The strong associations between the single slider and all three SSQ subscales may partly reflect participants' experience of symptoms beyond those specified during training and their use of the single available slider to report these symptoms. Finally, the slider itself may have served as a static visual reference and potentially mitigated cybersickness \cite{cao2018visually}. This potential mitigating effect was not explicitly considered in the study and may have influenced symptom severity across conditions.

Our multimodal dataset provides several opportunities for future research. Because we collected physiological data before VR exposure, including ECG, GSR, PPG, and eye tracking, future work can investigate whether baseline physiological characteristics can predict which cybersickness symptoms a person may experience during VR. The availability of both outside-VR and in-VR baseline measurements also allows future studies to examine the physiological changes caused by entering the virtual environment and whether these early changes can predict later cybersickness. Time-series prediction methods could further be used to develop personalized models that detect symptoms early and potentially support preventive interventions \cite{long2025toward, zhu2025towards, dennison2016use}. The continuous symptom-specific measurements can also be used to evaluate targeted cybersickness interventions by examining whether a particular intervention reduces one symptom more effectively than another. Future work on this data could look further into how specific symptoms affect attention in VR, for example by separating each participant's average severity from their moment-to-moment change, or by accounting for time spent in the simulation. Finally, although using three sliders did not show a significant increase in cognitive load compared with one slider, future studies should investigate the maximum number of symptoms that users can reliably report simultaneously and determine how many input dimensions can be supported before the interface begins to interfere with attention or task performance.
\section{Conclusion}
This study introduced a novel three-slider interface for the continuous, multidimensional reporting of cybersickness in virtual reality. By moving beyond single-metric discomfort scores, this tool allows researchers and developers to track the onset and trajectory of three  symptoms—nausea, oculomotor, and disorientation. For the broader VR community, this granular level of symptom tracking is a critical stepping stone toward developing adaptive environments that can automatically adjust rendering, locomotion, or field-of-view in real time to mitigate specific discomforts before they escalate. 
 By validating this multidimensional reporting tool and openly sharing our comprehensive multimodal dataset, this work establishes a robust methodological foundation for the future of real-time cybersickness prediction and targeted symptom mitigation.

\section{Acknowledgments}

This study was reviewed and approved by the university's Institutional Review Board (University of Texas at San Antonio IRB, Protocol
\#14-095, approved 11 January 2024; contact irb@utsa.edu) before any data collection began. All participants gave written informed consent before the session and were explicitly told that the VR exposure could induce mild to severe cybersickness symptoms. They were informed that they could withdraw at any time, for any reason, without penalty, and could stop the simulation whenever they wished. They were instructed to notify the research team immediately if they experienced severe discomfort. Participants were compensated \$30 per hour.

\bibliographystyle{ACM-Reference-Format}
\bibliography{sample-base}
\clearpage
\appendix
\section{ Appendix: Wearable sensors}
\begin{table}[htbp]
\centering
\caption{Overview of Collected Data Modalities, Features, Sampling Rates, and Devices}
\label{tab:data_modalities}
\renewcommand{\arraystretch}{1.3}
\begin{tabular}{p{3cm} p{4cm} c p{3cm}}
\hline
\textbf{Modality} & \textbf{Features} & \textbf{Sampling Rate} & \textbf{Device} \\ \hline

GSR & Galvanic skin response & 128\,Hz & Shimmer Consensys GSR \\
PPG & Photoplethysmogram & 128\,Hz & Shimmer Consensys GSR \\
Accelerometer / Gyroscope & Motion & 128\,Hz & Shimmer Consensys GSR \\
ECG & Cardiac electrical activity & 120\,Hz & Polar H10 \\
Eye Tracking (baseline) & Gaze direction, pupil diameter & 200\,Hz & Pupil Core \\
Eye \& Head Tracking (simulation) & Gaze direction, pupil diameter, eye openness, head orientation & 120\,Hz & HP Omnicept SDK (Reverb G2) \\
\hline

\end{tabular}
\end{table}
\section{Appendix: Rationale for Baseline Data Collection}
\label{sec:appendix-baseline-rationale}

We collected pre-exposure baseline data to characterize each participant's
typical physiological and eye-movement state before exposure to the
motion-provoking VR simulation. Physiological measures such as ECG, GSR, and PPG vary substantially across individuals \cite{coste2025comparative}. The
baseline recording therefore allowed simulation-window features to be expressed
relative to each participant's own pre-exposure state, reducing the influence of between-participant differences.

For eye tracking baseline, participants viewed a uniform grey background while following
white spheres that appeared at different locations. The grey background provided
stable, visually simple illumination and avoided optic flow, scene motion, and
visually complex content before the simulation. This was particularly important
for pupil measures, because pupil diameter is strongly affected by the luminance
of the viewed region. Maintaining a consistent background and target appearance
reduced pupil-size changes caused by luminance variation rather than
cybersickness~\cite{mathot2018safe,Winn2018}.

The appearing spheres provided standardized gaze targets across multiple display
locations. Asking participants to follow the spheres allowed us to obtain
controlled samples of fixation and saccade behavior while limiting variation
from unconstrained free viewing. Fixation- and saccade-related measures have
been examined as potential indicators of cybersickness and cognitive demands in
VR~\cite{Lee2021PredictingCybersickness, Nasri2024CognitiveLoad}. Thus, the
eye-tracking baseline provided a participant-specific reference for these
measures before participants encountered visually provocative virtual motion.

We repeated the grey-background and sphere-following procedure inside the HMD.
This in-VR baseline captured eye-tracking data under the headset's display,
optical, and eye-tracking geometry, which can differ from conventional
screen-based viewing. Together, the outside-VR and in-VR baseline recordings
helped distinguish participant-specific pre-exposure characteristics from
changes associated with the VR simulation.
\section{Appendix: Data Preprocessing Details}
\label{sec:appendix-preprocessing}
\paragraph{Eye tracking} Raw gaze samples were first cleaned: samples flagged invalid by the eye tracker (e.g., during blinks or tracking loss) were excluded, and pupil diameter was linearly interpolated across those short gaps. Fixations and saccades were then classified from the cleaned gaze stream using an Identification by Velocity Threshold (I-VT) algorithm. From this classified stream we calculated fixation counts and mean fixation durations, blink counts and durations, saccade counts and rates, peak velocities and amplitudes, and overall gaze dispersion, together with pupil diameter (mean, standard deviation, and within-window slope).
 
\paragraph{Head tracking} Raw head-orientation samples (quaternions from the headset's inside-out tracking) were converted to Euler angles and unwrapped to remove discontinuities at the $\pm180^\circ$ boundary before feature computation. Head-tracking features summarized angular velocity (mean, standard deviation) and total rotational range from the unwrapped angle series.
 
\paragraph{Cardiac signals (ECG, PPG)} ECG and PPG signals were cleaned and beat-detected using NeuroKit2. Cleaning followed NeuroKit2's default pipeline for each signal type: ECG was high-pass filtered (Butterworth, 0.5~Hz cutoff) to remove baseline wander, with powerline interference removed via a notch filter; PPG was band-pass filtered (Butterworth, 0.5--8~Hz) to isolate the pulsatile component before peak detection. Inter-beat intervals were further filtered to a physiologically plausible 30--240~bpm range to reject artifactual beats before feature computation. From the resulting beat series we derived, per window, mean heart rate, beat count, and standard heart-rate-variability metrics: SDNN (the standard deviation of inter-beat intervals), RMSSD (the root-mean-square of successive inter-beat-interval differences), and, for ECG, pNN50 (the proportion of successive interval differences exceeding 50~ms). The same beat-detection and HRV pipeline was applied independently to the chest-strap ECG and the wrist/finger PPG signal, yielding parallel feature sets for both.
 
\paragraph{Electrodermal activity (GSR)} GSR was processed with NeuroKit2's EDA pipeline. The raw signal was first low-pass filtered (Butterworth, 3~Hz cutoff) to remove high-frequency noise, then decomposed into a tonic skin-conductance level (SCL) and a phasic component from which discrete skin-conductance responses (SCRs) are detected. From this decomposition we derived, per window, mean SCL, SCR count, SCR rate (per minute), and mean SCR amplitude, together with the standard deviation of the raw (unfiltered) GSR signal as a coarse measure of overall electrodermal variability.
\section{Appendix: Study Protocol} 
\label{app:study_protocol}
This study was reviewed and approved by the Institutional Review Board (IRB) of ANONYMOUS prior to data collection. All participants provided written informed consent before beginning the study and were informed that the VR experience could induce mild to severe cybersickness symptoms.

\textbf{Inclusion Criteria.} Participants were required to be at least 18 years of age, have normal or corrected-to-normal vision, and be able to attend the study session in person.

\textbf{Exclusion Criteria.} Participants were excluded if they reported a history of epilepsy, vestibular disorders, pregnancy or hearing impairments, as well as any other condition that could be exacerbated by VR exposure or interfere with the auditory Simon task. Participants with uncorrected vision impairments that could interfere with the VR display or eye-tracking calibration were also excluded. Participants were instructed to refrain from consuming a heavy meal for at least one hour prior to the study.
Participants were informed that they could withdraw from the study at any time, for any reason, without penalty, and were explicitly instructed to notify the research team immediately if they experienced severe discomfort during the simulation.

\section{Appendix: Description of Extracted Features}

This appendix provides a brief description of all the physiological and behavioral features extracted from the dataset over predefined windows.

\subsection*{Electrocardiogram (ECG) Features}
The ECG features primarily capture heart rate and heart-rate variability (HRV) metrics.
\begin{itemize}
    \item \textbf{ecg\_hr\_mean\_bpm}: Mean heart rate in beats per minute (BPM).
    \item \textbf{ecg\_hr\_std\_bpm}: Standard deviation of the heart rate.
    \item \textbf{ecg\_sdnn\_ms}: Standard deviation of normal-to-normal RR intervals in milliseconds.
    \item \textbf{ecg\_rmssd\_ms}: Root mean square of successive differences between normal RR intervals in milliseconds.
    \item \textbf{ecg\_pnn50}: Proportion of successive RR interval pairs that differ by more than 50 ms.
    \item \textbf{ecg\_beat\_count}: Total number of detected heartbeats within the window.
\end{itemize}

\subsection*{Photoplethysmography (PPG) Features}
The PPG features similarly capture heart rate, HRV metrics, and signal amplitude statistics.
\begin{itemize}
    \item \textbf{ppg\_hr\_mean\_bpm}: Mean heart rate calculated from PPG peaks.
    \item \textbf{ppg\_hr\_std\_bpm}: Standard deviation of the heart rate.
    \item \textbf{ppg\_sdnn\_ms}: Standard deviation of inter-beat intervals in milliseconds.
    \item \textbf{ppg\_rmssd\_ms}: Root mean square of successive differences between inter-beat intervals.
    \item \textbf{ppg\_amplitude\_mean}: Mean amplitude of the raw PPG signal.
    \item \textbf{ppg\_amplitude\_std}: Standard deviation of the raw PPG signal amplitude.
    \item \textbf{ppg\_beat\_count}: Total number of detected beats from the PPG signal.
\end{itemize}

\subsection*{Eye-Tracking Features}
Eye-tracking features are derived from pupil diameter, blink events, and I-VT (Velocity-Threshold Identification) based gaze fixations and saccades.
\begin{itemize}
    \item \textbf{eye\_pupil\_diam\_mean}: Mean combined pupil diameter.
    \item \textbf{eye\_pupil\_diam\_std}: Standard deviation of the pupil diameter.
    \item \textbf{eye\_pupil\_diam\_slope}: Linear regression slope of the pupil diameter over the window.
    \item \textbf{eye\_pupil\_valid\_frac}: Fraction of the window with valid (non-zero) pupil diameter readings.
    \item \textbf{eye\_blink\_rate\_per\_min}: Number of blinks per minute.
    \item \textbf{eye\_blink\_duration\_mean\_sec}: Mean duration of blinks in seconds.
    \item \textbf{eye\_fixation\_count}: Total number of gaze fixations.
    \item \textbf{eye\_fixation\_duration\_mean\_sec}: Mean duration of fixations in seconds.
    \item \textbf{eye\_fixation\_time\_frac}: Fraction of time spent in fixation over the window.
    \item \textbf{eye\_saccade\_count}: Total number of saccades.
    \item \textbf{eye\_saccade\_rate\_per\_min}: Rate of saccades per minute.
    \item \textbf{eye\_saccade\_peak\_vel\_mean\_deg\_s}: Mean peak angular velocity during saccades (degrees/second).
    \item \textbf{eye\_saccade\_amplitude\_mean\_deg}: Mean amplitude (distance) of saccades in degrees.
    \item \textbf{eye\_gaze\_dispersion\_deg}: Dispersion of the gaze vectors across the window in degrees.
\end{itemize}

\subsection*{Galvanic Skin Response (GSR) Features}
GSR features summarize electrodermal activity, specifically focusing on tonic levels (Skin Conductance Level - SCL) and phasic responses (Skin Conductance Responses - SCR).
\begin{itemize}
    \item \textbf{gsr\_scl\_mean}: Mean skin conductance level.
    \item \textbf{gsr\_scl\_slope}: Linear regression slope of the skin conductance level over time.
    \item \textbf{gsr\_raw\_std}: Standard deviation of the raw skin conductance signal.
    \item \textbf{gsr\_scr\_count}: Total number of skin conductance response peaks (bursts).
    \item \textbf{gsr\_scr\_rate\_per\_min}: Rate of SCR bursts per minute.
    \item \textbf{gsr\_scr\_amplitude\_mean}: Mean amplitude of the detected SCR peaks.
\end{itemize}

\subsection*{Head-Tracking Features}
Head-tracking features serve as proxies for postural instability and head movement, derived from Euler angles.
\begin{itemize}
    \item \textbf{head\_ang\_vel\_mean\_deg\_s}: Mean rotational angular velocity in degrees per second.
    \item \textbf{head\_ang\_vel\_std\_deg\_s}: Standard deviation of the rotational angular velocity.
    \item \textbf{head\_rot\_range\_deg}: Overall range of rotation based on unwrapped Euler angles across the three spatial axes.
\end{itemize}

\section{Appendix: Multimodal signal data representation}
\label{app:fig}
\begin{figure}[htbp]
    \centering
    \includegraphics[width=0.7\linewidth]{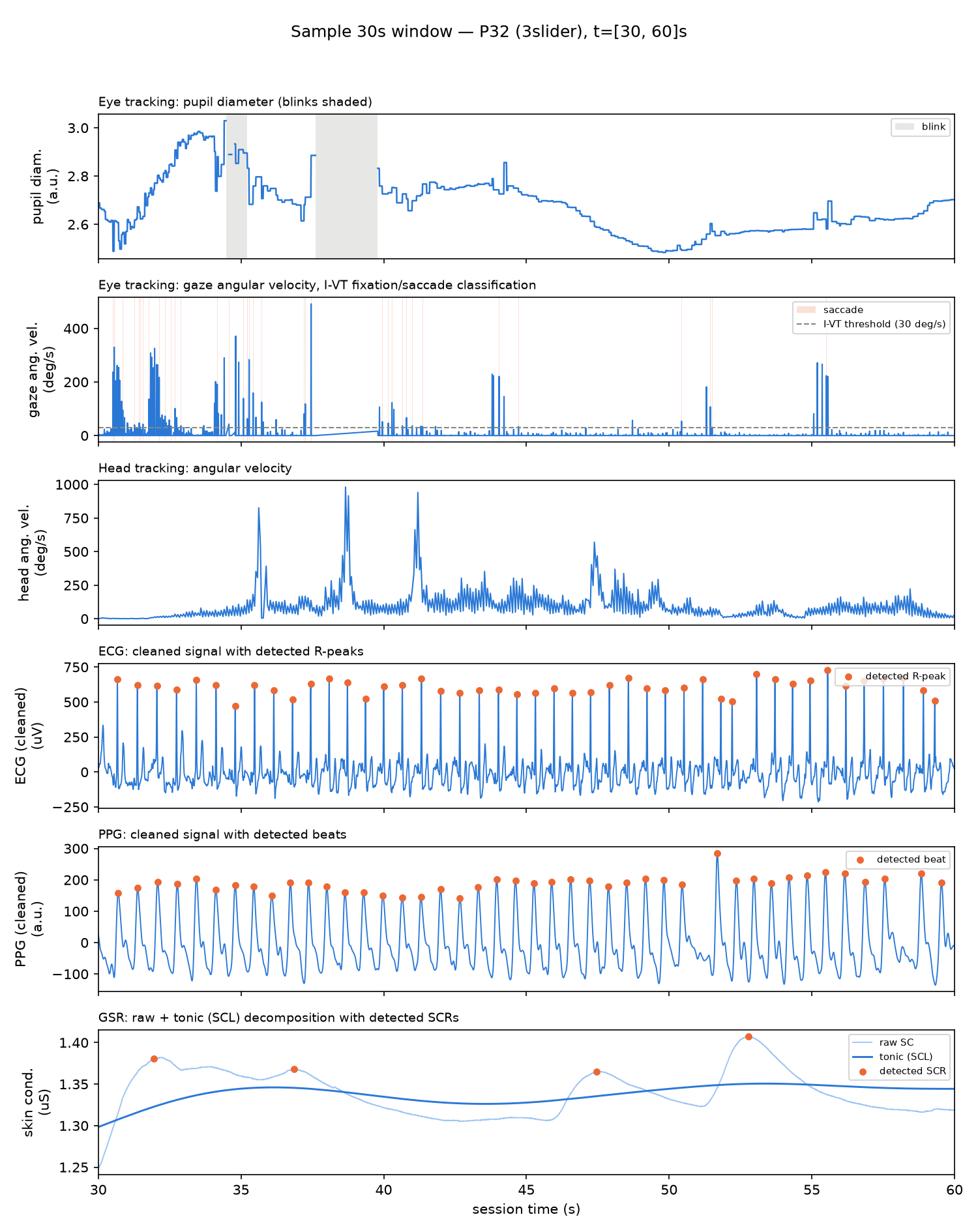}
    \caption{A representative 30~s window (session time 30--60~s) of processed multimodal data from a single participant (P32, 3-slider condition) (a) Pupil diameter, with blink periods shaded and excluded from feature computation. (b) Gaze angular velocity with the I-VT threshold (30~deg/s) used to classify saccades (shaded) from fixations. (c) Head angular velocity, showing several rapid head movements exceeding 500~deg/s. (d) Cleaned ECG signal with detected R-peaks. (e) Cleaned PPG signal with detected pulse beats. (f) Raw and tonic (SCL) galvanic skin response, with detected skin-conductance responses (SCRs) marked. Across modalities, event detection (saccades, R-peaks, PPG beats, SCRs) tracks the underlying signal closely, supporting the reliability of the windowed features derived from this pipeline.}
    \label{fig:sample-window}
\end{figure}
\end{document}